\documentclass[lettersize,journal]{IEEEtran}

\usepackage[normalem]{ulem}
\usepackage{amsmath,amsfonts}
\usepackage{algorithmic}
\usepackage{algorithm}
\usepackage{array}
\usepackage[caption=false,font=normalsize,labelfont=sf,textfont=sf]{subfig}
\usepackage{textcomp}
\usepackage{stfloats}
\usepackage{url}
\usepackage{verbatim}
\usepackage{graphicx}
\usepackage{cite}
\usepackage{nicefrac}
\usepackage[dvipsnames]{xcolor}
\usepackage[automark]{scrlayer-scrpage}
\setkomafont{pageheadfoot}{%
\normalfont\normalcolor\footnotesize}

\usepackage{ifthen} 
\newcommand{\IEEEversion}{false}        

\usepackage[hidelinks]{hyperref}
\hypersetup{
    colorlinks=false,
    linkcolor=black
    }

\usepackage{tcolorbox}
\usepackage{bm}			                
\usepackage[per-mode=symbol]{siunitx}   
\usepackage{cite} 
\DeclareSIUnit{\dBm}{dBm}	            
\DeclareSIUnit{\eq}{eq}	                
\DeclareSIUnit{\sqrtW}{\ensuremath{\sqrt{\text{W}}}}
\usepackage{layouts}
\usepackage{overpic}		            
\usepackage{array, booktabs, xltabular, multirow} 
\usepackage{amssymb}                    
\usepackage[capitalise]{cleveref}

\definecolor{IEEEblue}{RGB}{0 98 155}

\usepackage{adjustbox}          

\usepackage{enumerate}      
\usepackage[xindy]{glossaries}
\makenoidxglossaries%
\glsdisablehyper            
\newacronym{2d}{2D}{two-dimensional}
\newacronym{3d}{3D}{three-dimensional}
\newacronym{6g}{6G}{sixth generation}
\newacronym{adcs}{ADCs}{analog-to-digital converters}
\newacronym{adc}{ADC}{analog-to-digital converter}
\newacronym{aoa}{AOA}{angle-of-arrival}
\newacronym{aod}{AOD}{angle-of-departure}
\newacronym{asic}{ASIC}{application specific integrated circuit}
\newacronym{awgn}{AWGN}{additive white Gaussian noise}
\newacronym{arfh}{ARFH}{auxiliary RF harvester}
\newacronym{bibc}{BiBC}{bistatic backscatter communication}
\newacronym{bf}{BF}{beamformer}
\newacronym{ce}{CE}{channel estimation}
\newacronym{ced}{CED}{cumulative energy demand}
\newacronym{cdf}{CDF}{cumulative distribution function}
\newacronym{cjt}{CJT}{coherent joint transmission} 
\newacronym{csi}{CSI}{channel state information}
\newacronym{crlb}{CRLB}{Cram\'er-Rao lower bound}
\newacronym{dac}{DAC}{digital-to-analog converter}
\newacronym{dc}{DC}{direct current}
\newacronym{dm}{DM}{diffuse multipath}
\newacronym{dmc}{DMC}{diffuse multipath component}
\newacronym{dli}{DLI}{direct link interference}
\newacronym{eh}{EH}{energy harvesting}
\newacronym{eirp}{EIRP}{equivalent isotropically radiated power}
\newacronym{elaa}{ELAA}{extremely large aperture array}
\newacronym{em}{EM}{electromagnetic}
\newacronym{en}{EN}{energy-neutral}     
\newacronym{end}{END}{energy-neutral device}    
\newacronym{e2e}{E2E}{End-to-End}
\newacronym{epu}{EPU}{edge processing unit}
\newacronym{esd}{ESD}{electrostatic discharge}
\newacronym{fa}{FA}{federation anchor}
\newacronym{jcrs}{JCRS}{joint communication and radar sensing }
\newacronym{ic}{IC}{integrated circuit}
\newacronym{iot}{IoT}{Internet of Things}
\newacronym{id}{ID}{information decoding}
\newacronym{isac}{ISAC}{integrated sensing and communication}
\newacronym{ism}{ISM}{industrial, scientific and medical}
\newacronym{kpi}{KPI}{key performance indicator}
\newacronym{lca}{LCA}{life cycle assessment}
\newacronym{lis}{LIS}{large intelligent surface}
\newacronym{liion}{Li-ion}{lithium-ion}
\newacronym{los}{LoS}{line-of-sight}
\newacronym{lmo}{LMO}{lithium ion manganese oxide}
\newacronym{mc}{MC}{Monte Carlo}
\newacronym{mcu}{MCU}{microcontroller unit}
\newacronym{mf}{MF}{matched filter}
\newacronym{mmse}{MMSE}{minimum mean square error}
\newacronym{mmwave}{mmWave}{millimeter wave}
\newacronym{mobc}{MoBC}{monostatic backscatter communication}
\newacronym{mimo}{MIMO}{multiple-input multiple-output}
\newacronym{mmimo}{mMIMO}{massive MIMO}
\newacronym{miso}{MISO}{multiple-input single-output}
\newacronym{nlos}{NLoS}{non-LoS}
\newacronym{olos}{OLoS}{obstructed LoS}
\newacronym{xlmimo}{XL-MIMO}{extremely large-scale \acrshort{mimo}}
\newacronym{mrc}{MRC}{maximum ratio combining}
\newacronym{mrfh}{MRFH}{main RF harvester}
\newacronym{mrt}{MRT}{maximum ratio transmission}
\newacronym{ofdm}{OFDM}{orthogonal frequency-division multiplexing}
\newacronym{ota}{OTA}{over-the-air}
\newacronym{nb}{NB}{narrowband}
\newacronym{nr}{NR}{New Radio}
\newacronym{pa}{PA}{power amplifier}
\newacronym{pdp}{PDP}{power delay profile}
\newacronym{per}{PER}{packet error rate}
\newacronym{pl}{PL}{path loss}
\newacronym{pg}{PG}{path gain}
\newacronym{pc}{PC}{pilot count}
\newacronym{pw}{PW}{planar wavefront}
\newacronym{rcs}{RCS}{radar cross-section}
\newacronym{re}{RE}{radio element}
\newacronym{rf}{RF}{radio frequency}
\newacronym{rfid}{RFID}{\acrshort{rf} identification}
\newacronym{ris}{RIS}{reflective intelligent surface}
\newacronym{rms}{RMS}{root mean square}
\newacronym{rss}{RSS}{received signal strength}
\newacronym{rssi}{RSSI}{received signal strength indicator}
\newacronym{rw}{RW}{RadioWeave}
\newacronym{sa}{SA}{synthetic aperture}
\newacronym{sdg}{SDG}{Sustainable Development Goal}
\newacronym{sdn}{SDN}{software-defined network}
\newacronym{si}{SI}{self-interference}
\newacronym{sinr}{SINR}{signal-to-interference-plus-noise ratio}
\newacronym{siso}{SISO}{single-input single-output}
\newacronym{slam}{SLAM}{simultaneous localization and mapping}
\newacronym{smc}{SMC}{specular multipath component}
\newacronym{snr}{SNR}{signal-to-noise ratio}
\newacronym{s-parameter}{S-parameter}{scattering parameter}
\newacronym{svd}{SVD}{singular value decomposition}
\newacronym{sw}{SW}{spherical wavefront}
\newacronym{swipt}{SWIPT}{simultaneous wireless information and power transfer}
\newacronym{tdd}{TDD}{time-division duplexing}
\newacronym{tdoa}{TDOA}{time-difference-of-arrival}
\newacronym{toa}{TOA}{time-of-arrival}
\newacronym{tosm}{TOSM}{through–open–short–match}
\newacronym{ue}{UE}{user equipment}
\newacronym{uhf}{UHF}{ultra high frequency}
\newacronym{ula}{ULA}{uniform linear array}
\newacronym{upa}{UPA}{uniform planar array}
\newacronym{ura}{URA}{uniform rectangular array}
\newacronym{uwb}{UWB}{ultrawideband}
\newacronym{vna}{VNA}{vector network analyzer}
\newacronym{wb}{WB}{wideband}
\newacronym{wpt}{WPT}{wireless power transfer}
\newacronym{wsn}{WSN}{wireless sensor network}
\newacronym{xets}{XETS}{cross exponentially tapered slot}
\newacronym{zf}{ZF}{zero forcing}
\newacronym{csp}{CSP}{contact service point}
\newacronym{ecsp}{ECSP}{edge computing service point}
\newacronym{ap}{AP}{access point}
\usepackage{balance}
\usepackage{flushend}
\usepackage{multirow}

\tcbset{shield externalize}

\newcommand{\externalizeFigures}{false}
\ifthenelse{\equal{\externalizeFigures}{true}}
{
  \usepackage{pgfplots}
  \pgfplotsset{compat=newest}
  \usepgfplotslibrary{external}
  \tikzexternalize[prefix=externalized/]
  \usepackage{shellesc}
}
{}

\definecolor{IEEElightblue}{RGB}{0 181 226}
\definecolor{IEEEturquoise}{RGB}{0 156 166}
\definecolor{IEEEred}{RGB}{186 12 47}
\definecolor{IEEEgreen}{RGB}{0 132 61}
\definecolor{IEEElightgreen}{RGB}{120 190 32}
\definecolor{IEEEorange}{RGB}{225 163 0}
\definecolor{IEEEyellow}{RGB}{255 209 0}
\definecolor{IEEEviolett}{RGB}{152 29 151}
\definecolor{IEEEdarkmaroon}{RGB}{134 31 65}

\definecolor{mygray}{gray}{0.6}
\definecolor{RDlightgreen}{RGB}{141 192 69}
\definecolor{RDgreen}{RGB}{93 109 68}

\usepackage{pgfplots}
\pgfplotsset{compat=newest}
\usetikzlibrary{plotmarks}
\usetikzlibrary{arrows.meta}
\usetikzlibrary{shapes}
\usetikzlibrary{calc}
\usepackage{tikzscale}		        
\usetikzlibrary{fit,backgrounds}    
    \pgfdeclarelayer{behindBackground}
    \pgfdeclarelayer{background}
    \pgfdeclarelayer{foreground}
    \pgfsetlayers{behindBackground,background,main,foreground} 

\usepackage{orcidlink}

\newcommand{\mrm}[1]{ \mathrm{#1} }

\newlength\figureheight
\newlength\figurewidth 

\newcounter{romanNumerals}
\makeatletter
\def\bstctlcite{\@ifnextchar[{\@bstctlcite}{\@bstctlcite[@auxout]}}
\def\@bstctlcite[#1]#2{\@bsphack
  \@for\@citeb:=#2\do{%
    \edef\@citeb{\expandafter\@firstofone\@citeb}%
    \if@filesw
      \immediate\write\csname #1\endcsname{\string\citation{\@citeb}}%
    \fi
  }%
  \@esphack}
\makeatother

\usepackage[
    type={CC},
    modifier={by},
    version={4.0},
]{doclicense}

\usepackage{tikzpagenodes}   
\AddToHook{shipout/firstpage}{
  \begin{tikzpicture}[remember picture,overlay]
    \node[anchor=south west,
          xshift=0.6in,     
          yshift=5mm]      
      at (current page.south west)
      {\doclicenseImage[imagewidth=5em]} ;
  \end{tikzpicture}%
}

\begin{document}

\bstctlcite{IEEEexample:BSTcontrol} 

\title{\resizebox{\textwidth}{!}{Physically Large Apertures for Wireless Power Transfer:} \\
 {Performance and Regulatory Aspects}
 }

\author{IEEE Publication Technology,~\IEEEmembership{Staff,~IEEE,}
\author{\IEEEauthorblockN{%
Benjamin J.\,B. Deutschmann\,\orcidlink{0000-0002-2647-7662}\IEEEauthorrefmark{4},
Ulrich Muehlmann\,\orcidlink{0000-0003-1337-0054}\IEEEauthorrefmark{3},
Ahmet Kaplan\,\orcidlink{0000-0003-2287-1651}\IEEEauthorrefmark{2},
Gilles Callebaut\,\orcidlink{0000-0003-2413-986X}\IEEEauthorrefmark{1},
Thomas Wilding\,\orcidlink{0000-0001-9392-4269}\IEEEauthorrefmark{6},
Bert Cox\,\orcidlink{0000-0002-0049-6435}\IEEEauthorrefmark{1},
Liesbet Van der Perre\,\orcidlink{0000-0002-9158-9628}\IEEEauthorrefmark{1},
Fredrik Tufvesson\,\orcidlink{0000-0003-1072-0784}\IEEEauthorrefmark{7},
Erik G. Larsson\,\orcidlink{0000-0002-7599-4367}\IEEEauthorrefmark{2}, 
Klaus Witrisal\,\orcidlink{0000-0002-1056-2133}\IEEEauthorrefmark{4}
}

\thanks{The project has received funding from the European Union’s Horizon 2020 research and innovation program under grant agreement No 101013425 and from the European Union's Horizon Europe research and innovation program under grant agreement No~101192113.
}
\IEEEauthorblockA{\IEEEauthorrefmark{4}
Graz University of Technology, Austria}, 
\IEEEauthorblockA{\IEEEauthorrefmark{2}
Link\"oping University, Sweden}, 
\IEEEauthorblockA{\IEEEauthorrefmark{1}
KU Leuven, Belgium}\\
\IEEEauthorblockA{\IEEEauthorrefmark{3}
NXP Semiconductors, Austria}, 
\IEEEauthorblockA{\IEEEauthorrefmark{7}
Lund University, Sweden},
\IEEEauthorblockA{\IEEEauthorrefmark{6}%
Austrian Institute of Technology, Austria}
}
}

\markboth{July~2025}%
{July~2025}

\maketitle

\begin{abstract}
\Gls{wpt} is a promising service for the \gls{iot}, providing a cost-effective and sustainable solution to deploy so-called energy-neutral devices on a massive scale. 
The power received at the device side from a conventional transmit antenna with a physically small aperture decays rapidly with the distance. 
New opportunities arise from the transition from conventional far-field beamforming to near-field beam focusing. 
We argue that a \emph{physically large} aperture, that is large with respect to the distance to the receiver, enables a power budget that remains practically independent of distance. 
Distance-dependent array gain patterns allow focusing the power density maximum precisely at the device location, while reducing the power density near the infrastructure.
Physical aperture size is a key resource in enabling efficient yet regulatory-compliant \gls{wpt}.
We use real-world measurements to demonstrate that a regulatory-compliant system operating at sub-10GHz frequencies can increase the power received at the device into the milliwatt range. 
Our empirical demonstration shows that power-optimal near-field beam focusing inherently exploits multipath propagation, yielding both increased WPT efficiency and improved human exposure safety. 
\end{abstract}


\glsresetall

\section{Introduction}
The use of the \gls{iot} is expected to grow exponentially in applications such as healthcare, logistics, and smart cities.
This brings significant sustainability challenges, particularly regarding the ecological impact of electronics manufacturing, and the ecotoxicity of batteries \cite{Hamed23IoT}.
\Gls{en} devices, which solely harvest ambient energy and run virtually indefinitely without batteries, tackle these issues by miniaturizing circuits, reducing waste, and extending device lifespans~\cite{Lopez22RadioStripesWPT,Clerckx21WPT}.
\Gls{rf} \gls{wpt} technology then powers these devices, but requires an infrastructure capable of providing efficient \gls{wpt} as a service on a massive scale.
Key additional challenges are
substantial infrastructure complexity and cost, limited receive powers and efficiencies, as well as stringent regulatory limits.

In this article, we show that physically large apertures --- meaning apertures with dimensions that are in a similar magnitude order as the propagation distances of interest --- are advantageous for \gls{wpt}.
Specifically, we explain why the use of large apertures results in improved \gls{wpt} efficiency, namely the \gls{pg},\footnote{The \gls{pg} is defined as the ratio of power received at the device to total power radiated by all transmit antennas.} and in lower power densities close to the array -- which is important in order to stay within regulatory limits on human exposure.
Physically large apertures naturally arise with the large and distributed antenna arrays that are envisioned for \gls{6g} systems, when operating at ``low'' carrier frequencies, especially in the ``golden'' bands below
\SI{10}{\giga\hertz}~\cite{Bjoernson24giganticMIMO}. 

\def\datapath{./figures}
\begin{figure}
        \setlength{\figurewidth}{0.95\columnwidth}
        \input{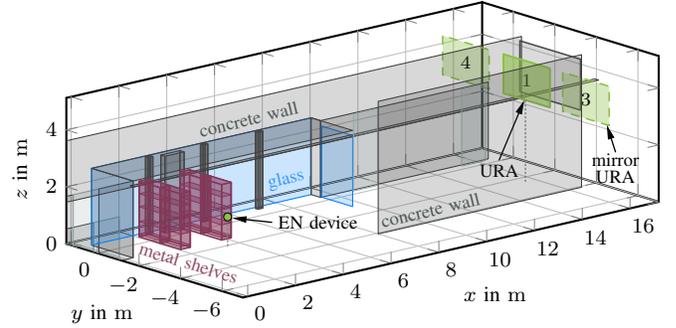}
        \caption{The hallway measurement scenario detailed in~\cite[Sec.\,V]{Deutschmann23ICC}: The specular multipath channel is modeled through an image source model. 
        Mirror arrays model walls. The mirror source $k\!=\!2$ is located below ground level.
        }
        \label{fig:MagazineScenario}
\end{figure}

To appreciate the main phenomenology, we consider an introductory example comparing two systems. 
The first system uses a physically large $(40\times 25)$-\gls{ura} of \nicefrac{3}{4}-wavelength spaced\footnote{No grating lobes appear at the considered steering angles.} antennas, 
operating at \SI{3.8}{\giga\hertz} (see Fig.\,\ref{fig:MagazineScenario}), and the second system uses a physically small $(40\times 25)$-\gls{ura} of \nicefrac{3}{4}-wavelength spaced antennas operating at \SI{38}{\giga\hertz}.
In both systems the array radiates \SI{10}{\watt} of total power and beamforms power to an \gls{en} device located \SI{12.3}{\metre} away from the array, in its boresight direction. 
Conjugate beamforming (maximum-ratio transmission)~\cite{mMIMObook} is used, assuming perfect \gls{csi} at the array. 
Importantly, such beamforming inherently accounts for near-field effects, that is, the fact that the wavefronts are curved.

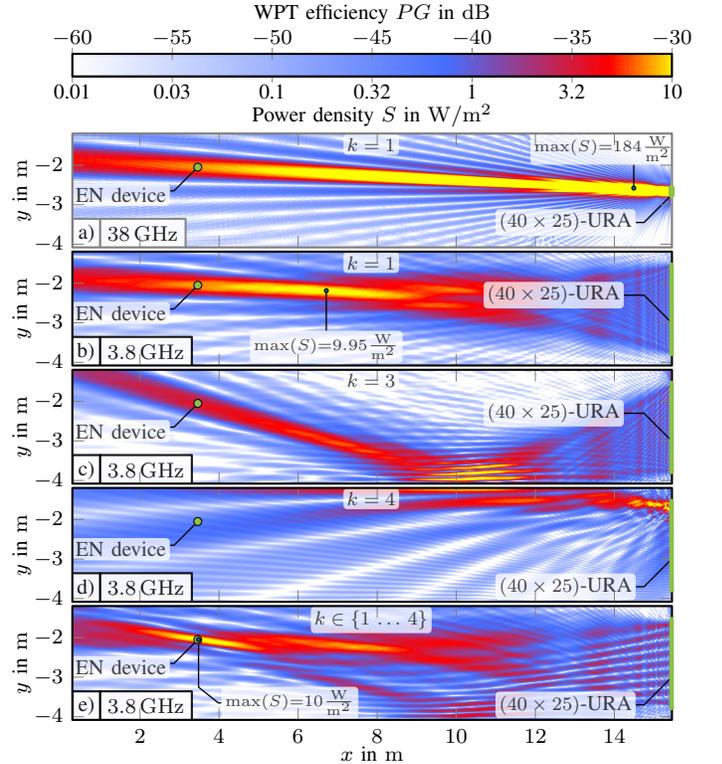
\begin{figure}[t]
        \def\datapath{./figures}
        \setlength{\figurewidth}{0.90\columnwidth}
        \hspace{-20mm}
%
%
\definecolor{mycolor1}{rgb}{0.55294,0.75294,0.27059}%
\definecolor{mygraylocal}{rgb}{0.5,0.5,0.5}%

\pgfplotsset{every axis/.append style={
  label style={font=\footnotesize},
  legend style={font=\footnotesize},
  tick label style={font=\footnotesize},
}}

\definecolor{IEEEblue}{RGB}{0 98 155}

\begin{tikzpicture}
\begin{axis}[
    width=\figurewidth,
    height=0.0001\figurewidth,
    at={(0\figurewidth, 0\figurewidth)},  
    scale only axis,
    line cap = round,
    line join = round,
    xmin=-60,
    xmax=-30,
    xtick={-60,-55,-50,-45,-40,-35,-30},                   
    xticklabels={0.01,0.03,0.1,0.32,1,3.2,10},  
    xtick pos=upper, %
    tick align=center, %
    xticklabel style = {yshift=0.1cm},
    ymin=1e-3,
    ymax=0,
    axis y line=none,
    xlabel={Power density $S$ in \SI{}{\watt\per\square\metre}},
    xlabel style = {yshift=0.17cm}
]
\end{axis}%
\begin{axis}[%
width=\figurewidth,
height=0.19\figurewidth,
at={(0\figurewidth,-0.285\figurewidth)},
scale only axis,
point meta min=-60,
point meta max=-30,
axis line style = {thick,mygraylocal},	
axis on top,
line cap = round,
line join = round,
xmin=0.29,
xmax=15.4559999999999,
xticklabel=\empty,                      
ymin=-4.08,
ymax=-1.2,
ylabel={$y$ in \SI{}{\metre}},        
ylabel style={yshift=-2mm},        
axis background/.style={fill=white},
colormap={mymap}{[1pt] rgb(0pt)=(0.996221,0.996221,0.996221); rgb(36pt)=(0.57167,0.663465,0.996221); rgb(37pt)=(0.559877,0.654221,0.996221); rgb(38pt)=(0.548083,0.644978,0.996221); rgb(39pt)=(0.53629,0.635735,0.996221); rgb(40pt)=(0.524497,0.626492,0.996221); rgb(41pt)=(0.512704,0.617248,0.996221); rgb(42pt)=(0.500911,0.608005,0.996221); rgb(43pt)=(0.489118,0.598762,0.996221); rgb(44pt)=(0.477325,0.589519,0.996221); rgb(45pt)=(0.465532,0.580275,0.996221); rgb(46pt)=(0.453739,0.571032,0.996221); rgb(47pt)=(0.441945,0.561789,0.996221); rgb(48pt)=(0.430152,0.552546,0.996222); rgb(49pt)=(0.418359,0.543302,0.996221); rgb(50pt)=(0.406567,0.534059,0.996221); rgb(51pt)=(0.394771,0.524816,0.996223); rgb(52pt)=(0.382981,0.515573,0.99622); rgb(53pt)=(0.371194,0.506328,0.996215); rgb(54pt)=(0.359374,0.49709,0.99624); rgb(55pt)=(0.347601,0.487843,0.996221); rgb(56pt)=(0.335907,0.478582,0.996126); rgb(57pt)=(0.323811,0.469392,0.996416); rgb(58pt)=(0.312052,0.460143,0.996384); rgb(59pt)=(0.301781,0.450632,0.994923); rgb(60pt)=(0.286814,0.441948,0.997969); rgb(61pt)=(0.272906,0.433077,1); rgb(62pt)=(0.287748,0.419142,0.974432); rgb(63pt)=(0.320483,0.402054,0.931687); rgb(64pt)=(0.347665,0.385945,0.894273); rgb(65pt)=(0.37359,0.370058,0.858067); rgb(66pt)=(0.401016,0.353906,0.820418); rgb(67pt)=(0.428144,0.337806,0.783056); rgb(68pt)=(0.455101,0.321737,0.745858); rgb(69pt)=(0.482176,0.305647,0.708547); rgb(70pt)=(0.509242,0.289558,0.671245); rgb(71pt)=(0.536289,0.273473,0.63396); rgb(72pt)=(0.563344,0.257386,0.596668); rgb(73pt)=(0.5904,0.241299,0.559375); rgb(74pt)=(0.617455,0.225213,0.522084); rgb(75pt)=(0.644509,0.209126,0.484792); rgb(76pt)=(0.671564,0.19304,0.447501); rgb(77pt)=(0.698619,0.176953,0.410209); rgb(78pt)=(0.725675,0.160861,0.372915); rgb(79pt)=(0.752726,0.14479,0.335629); rgb(80pt)=(0.779781,0.128702,0.298337); rgb(81pt)=(0.806856,0.112535,0.261017); rgb(82pt)=(0.833868,0.0966159,0.223784); rgb(83pt)=(0.860895,0.080641,0.186531); rgb(84pt)=(0.88823,0.0634504,0.148854); rgb(85pt)=(0.9149,0.0488806,0.112092); rgb(86pt)=(0.941208,0.0357377,0.0758298); rgb(87pt)=(0.971794,0.00572876,0.03367); rgb(88pt)=(0.996221,0,0); rgb(89pt)=(0.998849,0.0802077,0.0036217); rgb(90pt)=(0.995317,0.184696,0.00124634); rgb(91pt)=(0.996064,0.272317,0.000217044); rgb(92pt)=(0.996449,0.361366,0.000313345); rgb(93pt)=(0.996169,0.453036,7.27391e-05); rgb(94pt)=(0.996197,0.54349,3.37298e-05); rgb(95pt)=(0.99624,0.633886,2.54196e-05); rgb(96pt)=(0.996219,0.724534,3.42187e-06); rgb(97pt)=(0.996217,0.815106,5.47498e-06); rgb(98pt)=(0.996226,0.905636,6.84373e-06); rgb(99pt)=(0.996221,0.996221,0)},
colorbar horizontal,
colorbar style={%
at={(0,1.7)}, %
anchor=north west, %
height=0.3cm,%
xtick pos=upper, %
xticklabel pos=upper,%
xticklabel style={yshift = -0.1cm},%
xtick distance={5}, %
tick align=center, %
axis line style = thick,	
yshift = 1cm, %
xlabel={WPT efficiency $PG$ in \SI{}{\dB}},%
line cap = round,
line join = round,
xlabel style={yshift=-1.5mm}}
]

\addplot [forget plot] graphics [xmin=0.28, xmax=15.4, ymin=-4.09, ymax=-1.19] {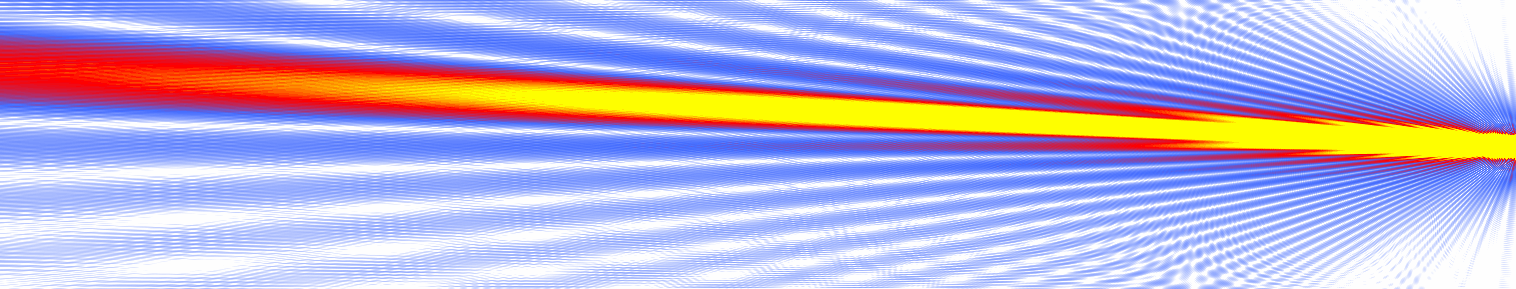};          
\addplot [color=mycolor1, only marks, mark size=0.7pt, mark=*, mark options={solid, mycolor1}, forget plot] 
  table[row sep=crcr]{%
15.4459999999999	-2.7770370401716\\
15.4459999999999	-2.7712915509320\\
15.4459999999999	-2.76554606169246\\
15.4459999999999	-2.75980057245289\\
15.4459999999999	-2.75405508321332\\
15.4459999999999	-2.74830959397375\\
15.4459999999999	-2.74256410473418\\
15.4459999999999	-2.73681861549461\\
15.4459999999999	-2.73107312625504\\
15.4459999999999	-2.72532763701548\\
15.4459999999999	-2.71958214777591\\
15.4459999999999	-2.71383665853634\\
15.4459999999999	-2.70809116929677\\
15.4459999999999	-2.7023456800572\\
15.4459999999999	-2.69660019081763\\
15.4459999999999	-2.69085470157806\\
15.4459999999999	-2.68510921233849\\
15.4459999999999	-2.67936372309892\\
15.4459999999999	-2.67361823385935\\
15.4459999999999	-2.66787274461978\\
15.4459999999999	-2.66212725538021\\
15.4459999999999	-2.65638176614065\\
15.4459999999999	-2.65063627690108\\
15.4459999999999	-2.64489078766151\\
15.4459999999999	-2.63914529842194\\
15.4459999999999	-2.63339980918237\\
15.4459999999999	-2.6276543199428\\
15.4459999999999	-2.62190883070323\\
15.4459999999999	-2.61616334146366\\
15.4459999999999	-2.61041785222409\\
15.4459999999999	-2.60467236298452\\
15.4459999999999	-2.59892687374495\\
15.4459999999999	-2.59318138450539\\
15.4459999999999	-2.58743589526582\\
15.4459999999999	-2.58169040602625\\
15.4459999999999	-2.57594491678668\\
15.4459999999999	-2.57019942754711\\
15.4459999999999	-2.56445393830754\\
15.4459999999999	-2.55870844906797\\
15.4459999999999	-2.5529629598284\\
};

\draw[color=black, line width=0.5pt]    
(15.45,-2.7500) -- %
(14.75,-3.4500) -- %
(14.55,-3.4500); %
\node[left, align=right ,font={\footnotesize},fill=white,
opacity=0.75,inner sep=0.5mm, xshift=0.0mm,rounded corners=0.5mm] at %
(axis cs:14.55,-3.4500){\scalebox{0.9}{$(40\times 25)$}-URA};

\draw[color=black, line width=0.5pt]    
(3.462,-2.050) -- %
(2.862,-2.650) -- %
(2.77,-2.650); %
\node[left, align=right ,font={\footnotesize},fill=white,
opacity=0.75,inner sep=0.5mm, xshift=0.0mm, rounded corners=0.5mm] at %
(axis cs:2.77,-2.750){EN device};

\addplot[only marks, mark=*, mark options={}, mark size=1.5000pt, color=black, fill=RDlightgreen, forget plot] table[row sep=crcr]{%
x	y\\
3.46166087184211	-2.05024565388158\\
}; 

\node[below, align=center ,font={\footnotesize},fill=white,
opacity=0.75,inner sep=0.5mm, xshift=0.0mm,rounded corners=0.5mm] at %
(axis cs:7.85,-1.2){\scalebox{0.9}{$k=1$}};  

\node[left, align=right ,font={\footnotesize}, draw, line width = 0.8pt, color=mygraylocal, fill=white, opacity=0.9,inner sep=0.75mm, xshift=0.5mm] at %
(axis cs:0.93,-3.73){\color{black}a)};

\node[right, align=left ,font={\footnotesize}, draw, line width = 0.8pt, color=mygraylocal, fill=white, opacity=0.9,inner sep=1.0mm, xshift=0.5mm,text=black] at %
(axis cs:0.87,-3.73){\SI{38}{\giga\hertz}};


\draw[color=black, line width=0.5pt]    
(14.49,-2.58) -- %
(14.49,-2.0); %
\node[above,align=center ,font={\footnotesize},fill=white,
opacity=0.75,inner sep=0.5mm, xshift=-4mm,xshift=0.0mm,rounded corners=0.5mm] at %
(axis cs:14.49,-2.0){$\scriptstyle\max(S)=184\frac{\SI{}{\watt}}{\SI{}{\square\metre}}$};

\addplot[only marks, mark=*, mark options={}, mark size=0.75000pt, color=black, fill=IEEEblue, forget plot] table[row sep=crcr]{%
x	y\\
14.49	-2.58\\
}; 

\end{axis}%
\end{tikzpicture}%
        \vspace{-6.2mm}%
        \hspace{-20mm}
%
%
\definecolor{mycolor1}{rgb}{0.55294,0.75294,0.27059}%

\pgfplotsset{every axis/.append style={
  label style={font=\footnotesize},
  legend style={font=\footnotesize},
  tick label style={font=\footnotesize},
}}

\begin{tikzpicture}

\begin{axis}[%
width=\figurewidth,
height=0.19\figurewidth,
at={(0\figurewidth,0\figurewidth)},
scale only axis,
point meta min=-60,
point meta max=-30,
axis line style = thick,	
line cap = round,
line join = round,
axis on top,
xmin=0.29,
xmax=15.4559999999999,
xticklabel=\empty,                      
ymin=-4.08,
ymax=-1.2,
ylabel={$y$ in \SI{}{\metre}},        
ylabel style={yshift=-2mm},        
axis background/.style={fill=white},
colormap={mymap}{[1pt] rgb(0pt)=(0.996221,0.996221,0.996221); rgb(36pt)=(0.57167,0.663465,0.996221); rgb(37pt)=(0.559877,0.654221,0.996221); rgb(38pt)=(0.548083,0.644978,0.996221); rgb(39pt)=(0.53629,0.635735,0.996221); rgb(40pt)=(0.524497,0.626492,0.996221); rgb(41pt)=(0.512704,0.617248,0.996221); rgb(42pt)=(0.500911,0.608005,0.996221); rgb(43pt)=(0.489118,0.598762,0.996221); rgb(44pt)=(0.477325,0.589519,0.996221); rgb(45pt)=(0.465532,0.580275,0.996221); rgb(46pt)=(0.453739,0.571032,0.996221); rgb(47pt)=(0.441945,0.561789,0.996221); rgb(48pt)=(0.430152,0.552546,0.996222); rgb(49pt)=(0.418359,0.543302,0.996221); rgb(50pt)=(0.406567,0.534059,0.996221); rgb(51pt)=(0.394771,0.524816,0.996223); rgb(52pt)=(0.382981,0.515573,0.99622); rgb(53pt)=(0.371194,0.506328,0.996215); rgb(54pt)=(0.359374,0.49709,0.99624); rgb(55pt)=(0.347601,0.487843,0.996221); rgb(56pt)=(0.335907,0.478582,0.996126); rgb(57pt)=(0.323811,0.469392,0.996416); rgb(58pt)=(0.312052,0.460143,0.996384); rgb(59pt)=(0.301781,0.450632,0.994923); rgb(60pt)=(0.286814,0.441948,0.997969); rgb(61pt)=(0.272906,0.433077,1); rgb(62pt)=(0.287748,0.419142,0.974432); rgb(63pt)=(0.320483,0.402054,0.931687); rgb(64pt)=(0.347665,0.385945,0.894273); rgb(65pt)=(0.37359,0.370058,0.858067); rgb(66pt)=(0.401016,0.353906,0.820418); rgb(67pt)=(0.428144,0.337806,0.783056); rgb(68pt)=(0.455101,0.321737,0.745858); rgb(69pt)=(0.482176,0.305647,0.708547); rgb(70pt)=(0.509242,0.289558,0.671245); rgb(71pt)=(0.536289,0.273473,0.63396); rgb(72pt)=(0.563344,0.257386,0.596668); rgb(73pt)=(0.5904,0.241299,0.559375); rgb(74pt)=(0.617455,0.225213,0.522084); rgb(75pt)=(0.644509,0.209126,0.484792); rgb(76pt)=(0.671564,0.19304,0.447501); rgb(77pt)=(0.698619,0.176953,0.410209); rgb(78pt)=(0.725675,0.160861,0.372915); rgb(79pt)=(0.752726,0.14479,0.335629); rgb(80pt)=(0.779781,0.128702,0.298337); rgb(81pt)=(0.806856,0.112535,0.261017); rgb(82pt)=(0.833868,0.0966159,0.223784); rgb(83pt)=(0.860895,0.080641,0.186531); rgb(84pt)=(0.88823,0.0634504,0.148854); rgb(85pt)=(0.9149,0.0488806,0.112092); rgb(86pt)=(0.941208,0.0357377,0.0758298); rgb(87pt)=(0.971794,0.00572876,0.03367); rgb(88pt)=(0.996221,0,0); rgb(89pt)=(0.998849,0.0802077,0.0036217); rgb(90pt)=(0.995317,0.184696,0.00124634); rgb(91pt)=(0.996064,0.272317,0.000217044); rgb(92pt)=(0.996449,0.361366,0.000313345); rgb(93pt)=(0.996169,0.453036,7.27391e-05); rgb(94pt)=(0.996197,0.54349,3.37298e-05); rgb(95pt)=(0.99624,0.633886,2.54196e-05); rgb(96pt)=(0.996219,0.724534,3.42187e-06); rgb(97pt)=(0.996217,0.815106,5.47498e-06); rgb(98pt)=(0.996226,0.905636,6.84373e-06); rgb(99pt)=(0.996221,0.996221,0)},
]
\addplot [forget plot] graphics [xmin=0.28, xmax=15.4, ymin=-4.09, ymax=-1.19] {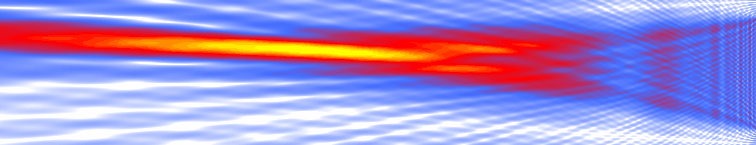};          
\addplot [color=mycolor1, only marks, mark size=0.7pt, mark=*, mark options={solid, mycolor1}, forget plot] 
  table[row sep=crcr]{%
15.446	-1.53580769544671\\
15.446	-1.59320217260329\\
15.446	-1.65059664975987\\
15.446	-1.70799112691645\\
15.446	-1.76538560407303\\
15.446	-1.82278008122961\\
15.446	-1.88017455838618\\
15.446	-1.93756903554276\\
15.446	-1.99496351269934\\
15.446	-2.05235798985592\\
15.446	-2.1097524670125\\
15.446	-2.16714694416908\\
15.446	-2.22454142132566\\
15.446	-2.28193589848224\\
15.446	-2.33933037563882\\
15.446	-2.3967248527954\\
15.446	-2.45411932995197\\
15.446	-2.51151380710855\\
15.446	-2.56890828426513\\
15.446	-2.62630276142171\\
15.446	-2.68369723857829\\
15.446	-2.74109171573487\\
15.446	-2.79848619289145\\
15.446	-2.85588067004803\\
15.446	-2.91327514720461\\
15.446	-2.97066962436118\\
15.446	-3.02806410151776\\
15.446	-3.08545857867434\\
15.446	-3.14285305583092\\
15.446	-3.2002475329875\\
15.446	-3.25764201014408\\
15.446	-3.31503648730066\\
15.446	-3.37243096445724\\
15.446	-3.42982544161382\\
15.446	-3.4872199187704\\
15.446	-3.54461439592697\\
15.446	-3.60200887308355\\
15.446	-3.65940335024013\\
15.446	-3.71679782739671\\
15.446	-3.77419230455329\\
15.446	-3.77419230455329\\
};

\draw[color=black, line width=0.5pt]    
(15.45,-3.000) -- %
(14.75,-2.300) -- %
(14.25,-2.300); %
\node[left, align=right ,font={\footnotesize},fill=white,
opacity=0.75,inner sep=0.5mm, xshift=0.0mm,rounded corners=0.5mm] at %
(axis cs:14.25,-2.300){\scalebox{0.9}{$(40\times 25)$}-URA};

\draw[color=black, line width=0.5pt]    
(3.462,-2.050) -- %
(2.862,-2.650) -- %
(2.77,-2.650); %
\node[left, align=right ,font={\footnotesize},fill=white,
opacity=0.75,inner sep=0.5mm, xshift=0.0mm, rounded corners=0.5mm] at %
(axis cs:2.77,-2.750){EN device};

\addplot[only marks, mark=*, mark options={}, mark size=1.5000pt, color=black, fill=RDlightgreen, forget plot] table[row sep=crcr]{%
x	y\\
3.46166087184211	-2.05024565388158\\
}; 

\node[below, align=center ,font={\footnotesize},fill=white,
opacity=0.75,inner sep=0.5mm, xshift=0.0mm,rounded corners=0.5mm] at %
(axis cs:7.85,-1.2){\scalebox{0.9}{$k=1$}};  

\node[left, align=right,font={\footnotesize},fill=white,
opacity=0.9,inner sep=0.75mm, xshift=0.5mm, draw, line width = 0.8pt] at %
(axis cs:0.93,-3.73){b)};

\node[right, align=left ,font={\footnotesize}, draw, line width = 0.8pt, fill=white, opacity=0.9,inner sep=1.0mm, xshift=0.5mm] at %
(axis cs:0.87,-3.73){\SI{3.8}{\giga\hertz}};

\draw[color=black, line width=0.5pt]    
(6.713,-3.1886) -- %
(6.713,-2.1886); %
\node[below,align=center ,font={\footnotesize},fill=white,
opacity=0.75,inner sep=0.5mm, xshift=0.0mm,rounded corners=0.5mm] at %
(axis cs:6.713,-3.1886){$\scriptstyle\max(S)=9.95\frac{\SI{}{\watt}}{\SI{}{\square\metre}}$};
\addplot[only marks, mark=*, mark options={}, mark size=0.75000pt, color=black, fill=IEEEblue, forget plot] table[row sep=crcr]{%
x	y\\
6.713	-2.1886\\
}; 

\end{axis}
\end{tikzpicture}%
        \vspace{-6.2mm}%
        \hspace{-20mm}
%
%
\definecolor{mycolor1}{rgb}{0.55294,0.75294,0.27059}%

\pgfplotsset{every axis/.append style={
  label style={font=\footnotesize},
  legend style={font=\footnotesize},
  tick label style={font=\footnotesize},
}}

\begin{tikzpicture}

\begin{axis}[%
width=\figurewidth,
height=0.19\figurewidth,
at={(0\figurewidth,0\figurewidth)},
scale only axis,
point meta min=-60,
point meta max=-30,
axis line style = thick,	
line cap = round,
line join = round,
axis on top,
xmin=0.29,
xmax=15.4559999999999,
xticklabel=\empty,                      
ymin=-4.08,
ymax=-1.2,
ylabel={$y$ in \SI{}{\metre}},          
ylabel style={yshift=-2mm},             
axis background/.style={fill=white},
colormap={mymap}{[1pt] rgb(0pt)=(0.996221,0.996221,0.996221); rgb(36pt)=(0.57167,0.663465,0.996221); rgb(37pt)=(0.559877,0.654221,0.996221); rgb(38pt)=(0.548083,0.644978,0.996221); rgb(39pt)=(0.53629,0.635735,0.996221); rgb(40pt)=(0.524497,0.626492,0.996221); rgb(41pt)=(0.512704,0.617248,0.996221); rgb(42pt)=(0.500911,0.608005,0.996221); rgb(43pt)=(0.489118,0.598762,0.996221); rgb(44pt)=(0.477325,0.589519,0.996221); rgb(45pt)=(0.465532,0.580275,0.996221); rgb(46pt)=(0.453739,0.571032,0.996221); rgb(47pt)=(0.441945,0.561789,0.996221); rgb(48pt)=(0.430152,0.552546,0.996222); rgb(49pt)=(0.418359,0.543302,0.996221); rgb(50pt)=(0.406567,0.534059,0.996221); rgb(51pt)=(0.394771,0.524816,0.996223); rgb(52pt)=(0.382981,0.515573,0.99622); rgb(53pt)=(0.371194,0.506328,0.996215); rgb(54pt)=(0.359374,0.49709,0.99624); rgb(55pt)=(0.347601,0.487843,0.996221); rgb(56pt)=(0.335907,0.478582,0.996126); rgb(57pt)=(0.323811,0.469392,0.996416); rgb(58pt)=(0.312052,0.460143,0.996384); rgb(59pt)=(0.301781,0.450632,0.994923); rgb(60pt)=(0.286814,0.441948,0.997969); rgb(61pt)=(0.272906,0.433077,1); rgb(62pt)=(0.287748,0.419142,0.974432); rgb(63pt)=(0.320483,0.402054,0.931687); rgb(64pt)=(0.347665,0.385945,0.894273); rgb(65pt)=(0.37359,0.370058,0.858067); rgb(66pt)=(0.401016,0.353906,0.820418); rgb(67pt)=(0.428144,0.337806,0.783056); rgb(68pt)=(0.455101,0.321737,0.745858); rgb(69pt)=(0.482176,0.305647,0.708547); rgb(70pt)=(0.509242,0.289558,0.671245); rgb(71pt)=(0.536289,0.273473,0.63396); rgb(72pt)=(0.563344,0.257386,0.596668); rgb(73pt)=(0.5904,0.241299,0.559375); rgb(74pt)=(0.617455,0.225213,0.522084); rgb(75pt)=(0.644509,0.209126,0.484792); rgb(76pt)=(0.671564,0.19304,0.447501); rgb(77pt)=(0.698619,0.176953,0.410209); rgb(78pt)=(0.725675,0.160861,0.372915); rgb(79pt)=(0.752726,0.14479,0.335629); rgb(80pt)=(0.779781,0.128702,0.298337); rgb(81pt)=(0.806856,0.112535,0.261017); rgb(82pt)=(0.833868,0.0966159,0.223784); rgb(83pt)=(0.860895,0.080641,0.186531); rgb(84pt)=(0.88823,0.0634504,0.148854); rgb(85pt)=(0.9149,0.0488806,0.112092); rgb(86pt)=(0.941208,0.0357377,0.0758298); rgb(87pt)=(0.971794,0.00572876,0.03367); rgb(88pt)=(0.996221,0,0); rgb(89pt)=(0.998849,0.0802077,0.0036217); rgb(90pt)=(0.995317,0.184696,0.00124634); rgb(91pt)=(0.996064,0.272317,0.000217044); rgb(92pt)=(0.996449,0.361366,0.000313345); rgb(93pt)=(0.996169,0.453036,7.27391e-05); rgb(94pt)=(0.996197,0.54349,3.37298e-05); rgb(95pt)=(0.99624,0.633886,2.54196e-05); rgb(96pt)=(0.996219,0.724534,3.42187e-06); rgb(97pt)=(0.996217,0.815106,5.47498e-06); rgb(98pt)=(0.996226,0.905636,6.84373e-06); rgb(99pt)=(0.996221,0.996221,0)},
]
\addplot [forget plot] graphics [xmin=0.28, xmax=15.4, ymin=-4.09, ymax=-1.19] {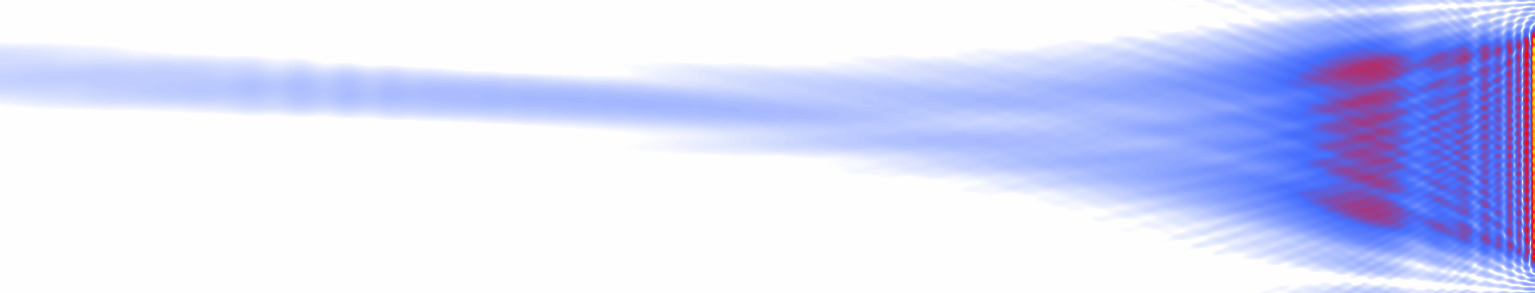};          
\addplot [color=mycolor1, only marks, mark size=0.7pt, mark=*, mark options={solid, mycolor1}, forget plot] 
  table[row sep=crcr]{%
15.446	-1.53580769544671\\
15.446	-1.59320217260329\\
15.446	-1.65059664975987\\
15.446	-1.70799112691645\\
15.446	-1.76538560407303\\
15.446	-1.82278008122961\\
15.446	-1.88017455838618\\
15.446	-1.93756903554276\\
15.446	-1.99496351269934\\
15.446	-2.05235798985592\\
15.446	-2.1097524670125\\
15.446	-2.16714694416908\\
15.446	-2.22454142132566\\
15.446	-2.28193589848224\\
15.446	-2.33933037563882\\
15.446	-2.3967248527954\\
15.446	-2.45411932995197\\
15.446	-2.51151380710855\\
15.446	-2.56890828426513\\
15.446	-2.62630276142171\\
15.446	-2.68369723857829\\
15.446	-2.74109171573487\\
15.446	-2.79848619289145\\
15.446	-2.85588067004803\\
15.446	-2.91327514720461\\
15.446	-2.97066962436118\\
15.446	-3.02806410151776\\
15.446	-3.08545857867434\\
15.446	-3.14285305583092\\
15.446	-3.2002475329875\\
15.446	-3.25764201014408\\
15.446	-3.31503648730066\\
15.446	-3.37243096445724\\
15.446	-3.42982544161382\\
15.446	-3.4872199187704\\
15.446	-3.54461439592697\\
15.446	-3.60200887308355\\
15.446	-3.65940335024013\\
15.446	-3.71679782739671\\
15.446	-3.77419230455329\\
15.446	-3.77419230455329\\
};

\draw[color=black, line width=0.5pt]    
(15.45,-3.000) -- %
(14.75,-2.300) -- %
(14.25,-2.300); %
\node[left, align=right ,font={\footnotesize},fill=white,
opacity=0.75,inner sep=0.5mm, xshift=0.0mm,rounded corners=0.5mm] at %
(axis cs:14.25,-2.300){\scalebox{0.9}{$(40\times 25)$}-URA};

\draw[color=black, line width=0.5pt]    
(3.462,-2.050) -- %
(2.862,-2.650) -- %
(2.77,-2.650); %
\node[left, align=right ,font={\footnotesize},fill=white,
opacity=0.75,inner sep=0.5mm, xshift=0.0mm, rounded corners=0.5mm] at %
(axis cs:2.77,-2.750){EN device};

\addplot[only marks, mark=*, mark options={}, mark size=1.5000pt, color=black, fill=RDlightgreen, forget plot] table[row sep=crcr]{%
x	y\\
3.46166087184211	-2.05024565388158\\
}; 

\node[below, align=center ,font={\footnotesize},fill=white,
opacity=0.75,inner sep=0.5mm, xshift=0.0mm,rounded corners=0.5mm] at %
(axis cs:7.85,-1.2){\scalebox{0.9}{$k=2$}};  

\node[left, align=right,font={\footnotesize},fill=white,
opacity=0.9,inner sep=0.75mm, xshift=0.5mm, draw, line width = 0.8pt] at %
(axis cs:0.93,-3.73){c)};

\node[right, align=left ,font={\footnotesize}, draw, line width = 0.8pt, fill=white, opacity=0.9,inner sep=1.0mm, xshift=0.5mm] at %
(axis cs:0.87,-3.73){\SI{3.8}{\giga\hertz}};

\end{axis}
\end{tikzpicture}%
        \vspace{-6.2mm}%
        \hspace{-20mm}
%
%
\definecolor{mycolor1}{rgb}{0.55294,0.75294,0.27059}%

\pgfplotsset{every axis/.append style={
  label style={font=\footnotesize},
  legend style={font=\footnotesize},
  tick label style={font=\footnotesize},
}}

\begin{tikzpicture}

\begin{axis}[%
width=\figurewidth,
height=0.19\figurewidth,
at={(0\figurewidth,0\figurewidth)},
scale only axis,
point meta min=-60,
point meta max=-30,
axis line style = thick,	
line cap = round,
line join = round,
axis on top,
xmin=0.29,
xmax=15.4559999999999,
xticklabel=\empty,                      
ymin=-4.08,
ymax=-1.2,
ylabel={$y$ in \SI{}{\metre}},          
ylabel style={yshift=-2mm},             
axis background/.style={fill=white},
colormap={mymap}{[1pt] rgb(0pt)=(0.996221,0.996221,0.996221); rgb(36pt)=(0.57167,0.663465,0.996221); rgb(37pt)=(0.559877,0.654221,0.996221); rgb(38pt)=(0.548083,0.644978,0.996221); rgb(39pt)=(0.53629,0.635735,0.996221); rgb(40pt)=(0.524497,0.626492,0.996221); rgb(41pt)=(0.512704,0.617248,0.996221); rgb(42pt)=(0.500911,0.608005,0.996221); rgb(43pt)=(0.489118,0.598762,0.996221); rgb(44pt)=(0.477325,0.589519,0.996221); rgb(45pt)=(0.465532,0.580275,0.996221); rgb(46pt)=(0.453739,0.571032,0.996221); rgb(47pt)=(0.441945,0.561789,0.996221); rgb(48pt)=(0.430152,0.552546,0.996222); rgb(49pt)=(0.418359,0.543302,0.996221); rgb(50pt)=(0.406567,0.534059,0.996221); rgb(51pt)=(0.394771,0.524816,0.996223); rgb(52pt)=(0.382981,0.515573,0.99622); rgb(53pt)=(0.371194,0.506328,0.996215); rgb(54pt)=(0.359374,0.49709,0.99624); rgb(55pt)=(0.347601,0.487843,0.996221); rgb(56pt)=(0.335907,0.478582,0.996126); rgb(57pt)=(0.323811,0.469392,0.996416); rgb(58pt)=(0.312052,0.460143,0.996384); rgb(59pt)=(0.301781,0.450632,0.994923); rgb(60pt)=(0.286814,0.441948,0.997969); rgb(61pt)=(0.272906,0.433077,1); rgb(62pt)=(0.287748,0.419142,0.974432); rgb(63pt)=(0.320483,0.402054,0.931687); rgb(64pt)=(0.347665,0.385945,0.894273); rgb(65pt)=(0.37359,0.370058,0.858067); rgb(66pt)=(0.401016,0.353906,0.820418); rgb(67pt)=(0.428144,0.337806,0.783056); rgb(68pt)=(0.455101,0.321737,0.745858); rgb(69pt)=(0.482176,0.305647,0.708547); rgb(70pt)=(0.509242,0.289558,0.671245); rgb(71pt)=(0.536289,0.273473,0.63396); rgb(72pt)=(0.563344,0.257386,0.596668); rgb(73pt)=(0.5904,0.241299,0.559375); rgb(74pt)=(0.617455,0.225213,0.522084); rgb(75pt)=(0.644509,0.209126,0.484792); rgb(76pt)=(0.671564,0.19304,0.447501); rgb(77pt)=(0.698619,0.176953,0.410209); rgb(78pt)=(0.725675,0.160861,0.372915); rgb(79pt)=(0.752726,0.14479,0.335629); rgb(80pt)=(0.779781,0.128702,0.298337); rgb(81pt)=(0.806856,0.112535,0.261017); rgb(82pt)=(0.833868,0.0966159,0.223784); rgb(83pt)=(0.860895,0.080641,0.186531); rgb(84pt)=(0.88823,0.0634504,0.148854); rgb(85pt)=(0.9149,0.0488806,0.112092); rgb(86pt)=(0.941208,0.0357377,0.0758298); rgb(87pt)=(0.971794,0.00572876,0.03367); rgb(88pt)=(0.996221,0,0); rgb(89pt)=(0.998849,0.0802077,0.0036217); rgb(90pt)=(0.995317,0.184696,0.00124634); rgb(91pt)=(0.996064,0.272317,0.000217044); rgb(92pt)=(0.996449,0.361366,0.000313345); rgb(93pt)=(0.996169,0.453036,7.27391e-05); rgb(94pt)=(0.996197,0.54349,3.37298e-05); rgb(95pt)=(0.99624,0.633886,2.54196e-05); rgb(96pt)=(0.996219,0.724534,3.42187e-06); rgb(97pt)=(0.996217,0.815106,5.47498e-06); rgb(98pt)=(0.996226,0.905636,6.84373e-06); rgb(99pt)=(0.996221,0.996221,0)},
]
\addplot [forget plot] graphics [xmin=0.28, xmax=15.4, ymin=-4.09, ymax=-1.19] {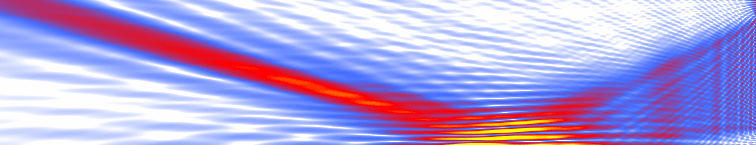};          
\addplot [color=mycolor1, only marks, mark size=0.7pt, mark=*, mark options={solid, mycolor1}, forget plot] 
  table[row sep=crcr]{%
15.446	-1.53580769544671\\
15.446	-1.59320217260329\\
15.446	-1.65059664975987\\
15.446	-1.70799112691645\\
15.446	-1.76538560407303\\
15.446	-1.82278008122961\\
15.446	-1.88017455838618\\
15.446	-1.93756903554276\\
15.446	-1.99496351269934\\
15.446	-2.05235798985592\\
15.446	-2.1097524670125\\
15.446	-2.16714694416908\\
15.446	-2.22454142132566\\
15.446	-2.28193589848224\\
15.446	-2.33933037563882\\
15.446	-2.3967248527954\\
15.446	-2.45411932995197\\
15.446	-2.51151380710855\\
15.446	-2.56890828426513\\
15.446	-2.62630276142171\\
15.446	-2.68369723857829\\
15.446	-2.74109171573487\\
15.446	-2.79848619289145\\
15.446	-2.85588067004803\\
15.446	-2.91327514720461\\
15.446	-2.97066962436118\\
15.446	-3.02806410151776\\
15.446	-3.08545857867434\\
15.446	-3.14285305583092\\
15.446	-3.2002475329875\\
15.446	-3.25764201014408\\
15.446	-3.31503648730066\\
15.446	-3.37243096445724\\
15.446	-3.42982544161382\\
15.446	-3.4872199187704\\
15.446	-3.54461439592697\\
15.446	-3.60200887308355\\
15.446	-3.65940335024013\\
15.446	-3.71679782739671\\
15.446	-3.77419230455329\\
15.446	-3.77419230455329\\
};

\draw[color=black, line width=0.5pt]    
(15.45,-3.000) -- %
(14.75,-2.300) -- %
(14.25,-2.300); %
\node[left, align=right ,font={\footnotesize},fill=white,
opacity=0.75,inner sep=0.5mm, xshift=0.0mm,rounded corners=0.5mm] at %
(axis cs:14.25,-2.300){\scalebox{0.9}{$(40\times 25)$}-URA};

\draw[color=black, line width=0.5pt]    
(3.462,-2.050) -- %
(2.862,-2.650) -- %
(2.77,-2.650); %
\node[left, align=right ,font={\footnotesize},fill=white,
opacity=0.75,inner sep=0.5mm, xshift=0.0mm, rounded corners=0.5mm] at %
(axis cs:2.77,-2.750){EN device};

\addplot[only marks, mark=*, mark options={}, mark size=1.5000pt, color=black, fill=RDlightgreen, forget plot] table[row sep=crcr]{%
x	y\\
3.46166087184211	-2.05024565388158\\
}; 

\node[below, align=center ,font={\footnotesize},fill=white,
opacity=0.75,inner sep=0.5mm, xshift=0.0mm,rounded corners=0.5mm] at %
(axis cs:7.85,-1.2){\scalebox{0.9}{$k=3$}};  

\node[left, align=right,font={\footnotesize},fill=white,
opacity=0.9,inner sep=0.75mm, xshift=0.5mm, draw, line width = 0.8pt] at %
(axis cs:0.93,-3.73){d)};

\node[right, align=left ,font={\footnotesize}, draw, line width = 0.8pt, fill=white, opacity=0.9,inner sep=1.0mm, xshift=0.5mm] at %
(axis cs:0.87,-3.73){\SI{3.8}{\giga\hertz}};

\end{axis}
\end{tikzpicture}%
        \vspace{-6.2mm}%
        \hspace{-20mm}
%
%
\definecolor{mycolor1}{rgb}{0.55294,0.75294,0.27059}%

\pgfplotsset{every axis/.append style={
  label style={font=\footnotesize},
  legend style={font=\footnotesize},
  tick label style={font=\footnotesize},
}}

\begin{tikzpicture}

\begin{axis}[%
width=\figurewidth,
height=0.19\figurewidth,
at={(0\figurewidth,0\figurewidth)},
scale only axis,
point meta min=-60,
point meta max=-30,
axis line style = thick,	
line cap = round,
line join = round,
axis on top,
xmin=0.29,
xmax=15.4559999999999,
xticklabel=\empty,                      
ymin=-4.08,
ymax=-1.2,
ylabel={$y$ in \SI{}{\metre}},        
ylabel style={yshift=-2mm},        
axis background/.style={fill=white},
colormap={mymap}{[1pt] rgb(0pt)=(0.996221,0.996221,0.996221); rgb(36pt)=(0.57167,0.663465,0.996221); rgb(37pt)=(0.559877,0.654221,0.996221); rgb(38pt)=(0.548083,0.644978,0.996221); rgb(39pt)=(0.53629,0.635735,0.996221); rgb(40pt)=(0.524497,0.626492,0.996221); rgb(41pt)=(0.512704,0.617248,0.996221); rgb(42pt)=(0.500911,0.608005,0.996221); rgb(43pt)=(0.489118,0.598762,0.996221); rgb(44pt)=(0.477325,0.589519,0.996221); rgb(45pt)=(0.465532,0.580275,0.996221); rgb(46pt)=(0.453739,0.571032,0.996221); rgb(47pt)=(0.441945,0.561789,0.996221); rgb(48pt)=(0.430152,0.552546,0.996222); rgb(49pt)=(0.418359,0.543302,0.996221); rgb(50pt)=(0.406567,0.534059,0.996221); rgb(51pt)=(0.394771,0.524816,0.996223); rgb(52pt)=(0.382981,0.515573,0.99622); rgb(53pt)=(0.371194,0.506328,0.996215); rgb(54pt)=(0.359374,0.49709,0.99624); rgb(55pt)=(0.347601,0.487843,0.996221); rgb(56pt)=(0.335907,0.478582,0.996126); rgb(57pt)=(0.323811,0.469392,0.996416); rgb(58pt)=(0.312052,0.460143,0.996384); rgb(59pt)=(0.301781,0.450632,0.994923); rgb(60pt)=(0.286814,0.441948,0.997969); rgb(61pt)=(0.272906,0.433077,1); rgb(62pt)=(0.287748,0.419142,0.974432); rgb(63pt)=(0.320483,0.402054,0.931687); rgb(64pt)=(0.347665,0.385945,0.894273); rgb(65pt)=(0.37359,0.370058,0.858067); rgb(66pt)=(0.401016,0.353906,0.820418); rgb(67pt)=(0.428144,0.337806,0.783056); rgb(68pt)=(0.455101,0.321737,0.745858); rgb(69pt)=(0.482176,0.305647,0.708547); rgb(70pt)=(0.509242,0.289558,0.671245); rgb(71pt)=(0.536289,0.273473,0.63396); rgb(72pt)=(0.563344,0.257386,0.596668); rgb(73pt)=(0.5904,0.241299,0.559375); rgb(74pt)=(0.617455,0.225213,0.522084); rgb(75pt)=(0.644509,0.209126,0.484792); rgb(76pt)=(0.671564,0.19304,0.447501); rgb(77pt)=(0.698619,0.176953,0.410209); rgb(78pt)=(0.725675,0.160861,0.372915); rgb(79pt)=(0.752726,0.14479,0.335629); rgb(80pt)=(0.779781,0.128702,0.298337); rgb(81pt)=(0.806856,0.112535,0.261017); rgb(82pt)=(0.833868,0.0966159,0.223784); rgb(83pt)=(0.860895,0.080641,0.186531); rgb(84pt)=(0.88823,0.0634504,0.148854); rgb(85pt)=(0.9149,0.0488806,0.112092); rgb(86pt)=(0.941208,0.0357377,0.0758298); rgb(87pt)=(0.971794,0.00572876,0.03367); rgb(88pt)=(0.996221,0,0); rgb(89pt)=(0.998849,0.0802077,0.0036217); rgb(90pt)=(0.995317,0.184696,0.00124634); rgb(91pt)=(0.996064,0.272317,0.000217044); rgb(92pt)=(0.996449,0.361366,0.000313345); rgb(93pt)=(0.996169,0.453036,7.27391e-05); rgb(94pt)=(0.996197,0.54349,3.37298e-05); rgb(95pt)=(0.99624,0.633886,2.54196e-05); rgb(96pt)=(0.996219,0.724534,3.42187e-06); rgb(97pt)=(0.996217,0.815106,5.47498e-06); rgb(98pt)=(0.996226,0.905636,6.84373e-06); rgb(99pt)=(0.996221,0.996221,0)},
]
\addplot [forget plot] graphics [xmin=0.28, xmax=15.4, ymin=-4.09, ymax=-1.19] {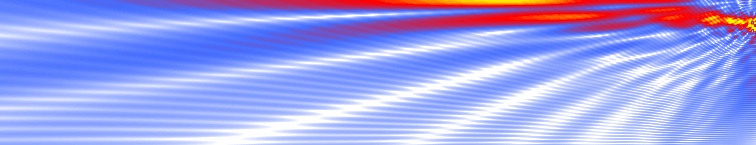};          
\addplot [color=mycolor1, only marks, mark size=0.7pt, mark=*, mark options={solid, mycolor1}, forget plot] 
  table[row sep=crcr]{%
15.446	-1.53580769544671\\
15.446	-1.59320217260329\\
15.446	-1.65059664975987\\
15.446	-1.70799112691645\\
15.446	-1.76538560407303\\
15.446	-1.82278008122961\\
15.446	-1.88017455838618\\
15.446	-1.93756903554276\\
15.446	-1.99496351269934\\
15.446	-2.05235798985592\\
15.446	-2.1097524670125\\
15.446	-2.16714694416908\\
15.446	-2.22454142132566\\
15.446	-2.28193589848224\\
15.446	-2.33933037563882\\
15.446	-2.3967248527954\\
15.446	-2.45411932995197\\
15.446	-2.51151380710855\\
15.446	-2.56890828426513\\
15.446	-2.62630276142171\\
15.446	-2.68369723857829\\
15.446	-2.74109171573487\\
15.446	-2.79848619289145\\
15.446	-2.85588067004803\\
15.446	-2.91327514720461\\
15.446	-2.97066962436118\\
15.446	-3.02806410151776\\
15.446	-3.08545857867434\\
15.446	-3.14285305583092\\
15.446	-3.2002475329875\\
15.446	-3.25764201014408\\
15.446	-3.31503648730066\\
15.446	-3.37243096445724\\
15.446	-3.42982544161382\\
15.446	-3.4872199187704\\
15.446	-3.54461439592697\\
15.446	-3.60200887308355\\
15.446	-3.65940335024013\\
15.446	-3.71679782739671\\
15.446	-3.77419230455329\\
15.446	-3.77419230455329\\
};

\draw[color=black, line width=0.5pt]    
(15.45,-3.000) -- %
(14.75,-3.700) -- %
(14.55,-3.700); %
\node[left, align=right ,font={\footnotesize},fill=white,
opacity=0.75,inner sep=0.5mm, xshift=0.0mm,rounded corners=0.5mm] at %
(axis cs:14.55,-3.700){\scalebox{0.9}{$(40\times 25)$}-URA};

\draw[color=black, line width=0.5pt]    
(3.462,-2.050) -- %
(2.862,-2.650) -- %
(2.77,-2.650); %
\node[left, align=right ,font={\footnotesize},fill=white,
opacity=0.75,inner sep=0.5mm, xshift=0.0mm, rounded corners=0.5mm] at %
(axis cs:2.77,-2.750){EN device};

\addplot[only marks, mark=*, mark options={}, mark size=1.5000pt, color=black, fill=RDlightgreen, forget plot] table[row sep=crcr]{%
x	y\\
3.46166087184211	-2.05024565388158\\
}; 

\node[below, align=center ,font={\footnotesize},fill=white,
opacity=0.75,inner sep=0.5mm, xshift=0.0mm,rounded corners=0.5mm] at %
(axis cs:7.85,-1.2){\scalebox{0.9}{$k=4$}};  

\node[left, align=right,font={\footnotesize},fill=white,
opacity=0.9,inner sep=0.75mm, xshift=0.5mm, draw, line width = 0.8pt] at %
(axis cs:0.93,-3.73){e)};

\node[right, align=left ,font={\footnotesize}, draw, line width = 0.8pt, fill=white, opacity=0.9,inner sep=1.0mm, xshift=0.5mm] at %
(axis cs:0.87,-3.73){\SI{3.8}{\giga\hertz}};

\end{axis}
\end{tikzpicture}%
        \vspace{-6.2mm}%
        \hspace{-20mm}
%
%
\definecolor{mycolor1}{rgb}{0.55294,0.75294,0.27059}%

\pgfplotsset{every axis/.append style={
  label style={font=\footnotesize},
  legend style={font=\footnotesize},
  tick label style={font=\footnotesize},
}}

\begin{tikzpicture}

\begin{axis}[%
width=\figurewidth,
height=0.19\figurewidth,
at={(0\figurewidth,0\figurewidth)},
scale only axis,
point meta min=-60,
point meta max=-30,
axis line style = thick,	
line cap = round,
line join = round,
axis on top,
xmin=0.29,
xmax=15.4559999999999,
xlabel={$x$ in \SI{}{\metre}},        
xlabel style={yshift=0.2cm},
ymin=-4.08,
ymax=-1.2,
ylabel={$y$ in \SI{}{\metre}},        
ylabel style={yshift=-2mm},        
axis background/.style={fill=white},
colormap={mymap}{[1pt] rgb(0pt)=(0.996221,0.996221,0.996221); rgb(36pt)=(0.57167,0.663465,0.996221); rgb(37pt)=(0.559877,0.654221,0.996221); rgb(38pt)=(0.548083,0.644978,0.996221); rgb(39pt)=(0.53629,0.635735,0.996221); rgb(40pt)=(0.524497,0.626492,0.996221); rgb(41pt)=(0.512704,0.617248,0.996221); rgb(42pt)=(0.500911,0.608005,0.996221); rgb(43pt)=(0.489118,0.598762,0.996221); rgb(44pt)=(0.477325,0.589519,0.996221); rgb(45pt)=(0.465532,0.580275,0.996221); rgb(46pt)=(0.453739,0.571032,0.996221); rgb(47pt)=(0.441945,0.561789,0.996221); rgb(48pt)=(0.430152,0.552546,0.996222); rgb(49pt)=(0.418359,0.543302,0.996221); rgb(50pt)=(0.406567,0.534059,0.996221); rgb(51pt)=(0.394771,0.524816,0.996223); rgb(52pt)=(0.382981,0.515573,0.99622); rgb(53pt)=(0.371194,0.506328,0.996215); rgb(54pt)=(0.359374,0.49709,0.99624); rgb(55pt)=(0.347601,0.487843,0.996221); rgb(56pt)=(0.335907,0.478582,0.996126); rgb(57pt)=(0.323811,0.469392,0.996416); rgb(58pt)=(0.312052,0.460143,0.996384); rgb(59pt)=(0.301781,0.450632,0.994923); rgb(60pt)=(0.286814,0.441948,0.997969); rgb(61pt)=(0.272906,0.433077,1); rgb(62pt)=(0.287748,0.419142,0.974432); rgb(63pt)=(0.320483,0.402054,0.931687); rgb(64pt)=(0.347665,0.385945,0.894273); rgb(65pt)=(0.37359,0.370058,0.858067); rgb(66pt)=(0.401016,0.353906,0.820418); rgb(67pt)=(0.428144,0.337806,0.783056); rgb(68pt)=(0.455101,0.321737,0.745858); rgb(69pt)=(0.482176,0.305647,0.708547); rgb(70pt)=(0.509242,0.289558,0.671245); rgb(71pt)=(0.536289,0.273473,0.63396); rgb(72pt)=(0.563344,0.257386,0.596668); rgb(73pt)=(0.5904,0.241299,0.559375); rgb(74pt)=(0.617455,0.225213,0.522084); rgb(75pt)=(0.644509,0.209126,0.484792); rgb(76pt)=(0.671564,0.19304,0.447501); rgb(77pt)=(0.698619,0.176953,0.410209); rgb(78pt)=(0.725675,0.160861,0.372915); rgb(79pt)=(0.752726,0.14479,0.335629); rgb(80pt)=(0.779781,0.128702,0.298337); rgb(81pt)=(0.806856,0.112535,0.261017); rgb(82pt)=(0.833868,0.0966159,0.223784); rgb(83pt)=(0.860895,0.080641,0.186531); rgb(84pt)=(0.88823,0.0634504,0.148854); rgb(85pt)=(0.9149,0.0488806,0.112092); rgb(86pt)=(0.941208,0.0357377,0.0758298); rgb(87pt)=(0.971794,0.00572876,0.03367); rgb(88pt)=(0.996221,0,0); rgb(89pt)=(0.998849,0.0802077,0.0036217); rgb(90pt)=(0.995317,0.184696,0.00124634); rgb(91pt)=(0.996064,0.272317,0.000217044); rgb(92pt)=(0.996449,0.361366,0.000313345); rgb(93pt)=(0.996169,0.453036,7.27391e-05); rgb(94pt)=(0.996197,0.54349,3.37298e-05); rgb(95pt)=(0.99624,0.633886,2.54196e-05); rgb(96pt)=(0.996219,0.724534,3.42187e-06); rgb(97pt)=(0.996217,0.815106,5.47498e-06); rgb(98pt)=(0.996226,0.905636,6.84373e-06); rgb(99pt)=(0.996221,0.996221,0)},
]
\addplot [forget plot] graphics [xmin=0.28, xmax=15.4, ymin=-4.09, ymax=-1.19] {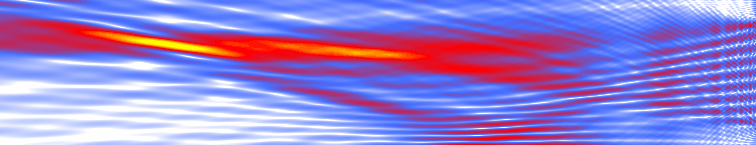};          
\addplot [color=mycolor1, only marks, mark size=0.7pt, mark=*, mark options={solid, mycolor1}, forget plot] 
  table[row sep=crcr]{%
15.446	-1.53580769544671\\
15.446	-1.59320217260329\\
15.446	-1.65059664975987\\
15.446	-1.70799112691645\\
15.446	-1.76538560407303\\
15.446	-1.82278008122961\\
15.446	-1.88017455838618\\
15.446	-1.93756903554276\\
15.446	-1.99496351269934\\
15.446	-2.05235798985592\\
15.446	-2.1097524670125\\
15.446	-2.16714694416908\\
15.446	-2.22454142132566\\
15.446	-2.28193589848224\\
15.446	-2.33933037563882\\
15.446	-2.3967248527954\\
15.446	-2.45411932995197\\
15.446	-2.51151380710855\\
15.446	-2.56890828426513\\
15.446	-2.62630276142171\\
15.446	-2.68369723857829\\
15.446	-2.74109171573487\\
15.446	-2.79848619289145\\
15.446	-2.85588067004803\\
15.446	-2.91327514720461\\
15.446	-2.97066962436118\\
15.446	-3.02806410151776\\
15.446	-3.08545857867434\\
15.446	-3.14285305583092\\
15.446	-3.2002475329875\\
15.446	-3.25764201014408\\
15.446	-3.31503648730066\\
15.446	-3.37243096445724\\
15.446	-3.42982544161382\\
15.446	-3.4872199187704\\
15.446	-3.54461439592697\\
15.446	-3.60200887308355\\
15.446	-3.65940335024013\\
15.446	-3.71679782739671\\
15.446	-3.77419230455329\\
15.446	-3.77419230455329\\
};

\draw[color=black, line width=0.5pt]    
(15.45,-3.000) -- %
(14.75,-3.700) -- %
(14.55,-3.700); %
\node[left, align=right ,font={\footnotesize},fill=white,
opacity=0.75,inner sep=0.5mm, xshift=0.0mm,rounded corners=0.5mm] at %
(axis cs:14.55,-3.700){\scalebox{0.9}{$(40\times 25)$}-URA};

\draw[color=black, line width=0.5pt]    
(3.462,-2.050) -- %
(2.862,-2.650) -- %
(2.77,-2.650); %
\node[left, align=right ,font={\footnotesize},fill=white,
opacity=0.75,inner sep=0.5mm, xshift=0.0mm, rounded corners=0.5mm] at %
(axis cs:2.77,-2.750){EN device};

\addplot[only marks, mark=*, mark options={}, mark size=1.5000pt, color=black, fill=RDlightgreen, forget plot] table[row sep=crcr]{%
x	y\\
3.46166087184211	-2.05024565388158\\
}; 

\node[below, align=center ,font={\footnotesize},fill=white,
opacity=0.75,inner sep=0.5mm, xshift=0.0mm,rounded corners=0.5mm] at %
(axis cs:7.85,-1.2){\scalebox{0.9}{$k\in\{1\,\dots\,4\}$}};  

\node[left, align=right,font={\footnotesize},fill=white,
opacity=0.9,inner sep=0.75mm, xshift=0.5mm, draw, line width = 0.8pt] at %
(axis cs:0.93,-3.73){f\,)};

\node[right, align=left ,font={\footnotesize}, draw, line width = 0.8pt, fill=white, opacity=0.9,inner sep=1.0mm, xshift=0.5mm] at %
(axis cs:0.87,-3.73){\SI{3.8}{\giga\hertz}};

\draw[color=black, line width=0.5pt]    
(4.05,-3.6506) --
(3.9818,-3.6506) -- %
(3.4818,-3.2506) -- %
(3.4818,-2.0506); %
\node[right,align=center ,font={\footnotesize},fill=white,
opacity=0.75,inner sep=0.5mm, xshift=0.0mm,rounded corners=0.5mm] at %
(axis cs:4.05,-3.6506){$\scriptstyle\max(S)=10\frac{\SI{}{\watt}}{\SI{}{\square\metre}}$};

\addplot[only marks, mark=*, mark options={}, mark size=0.75000pt, color=black, fill=IEEEblue, forget plot] table[row sep=crcr]{%
x	y\\
3.4818	-2.0506\\
}; 

\end{axis}
\end{tikzpicture}%
%
        \vspace{-3mm}%
        \caption{\acrshort{pg} simulated (using~\cite[eq.\,(2)]{Deutschmann23ICC}) on a plane connecting the \gls{ura} and the \gls{en} device for different SMCs. The power density scale assumes a total radiated power of \SI{10}{\watt}.
        Leveraging all multipath components $k\!\in\!\{1\,\dots\,4\}$ increases the efficiency and moves the global maximum of the power distribution to the \gls{en} device. The \SI{38}{\giga\hertz} simulation in a) is for comparison only.          }
        \label{fig:BFsim}
\end{figure}

Figures~\ref{fig:BFsim}\,a) and~\ref{fig:BFsim}\,b) show both the \acrlong{pg} $PG$ in \SI{}{\dB} and the power density $S$ in \SI{}{\watt\per\square\metre}, as a function of spatial location for these two systems, when using near-field beam focusing given the perfect \gls{los} channel as \gls{csi}.
We note the following:
\begin{itemize}
  \item For a given transmitted power, the power density at the \gls{en} device is the same in both systems.
    Hence, the harvested power (and \gls{wpt} efficiency) will be identical provided that the \gls{en} device has the same effective receive aperture (of $\SI{10}{\centi\metre\squared}$) both at \SI{3.8}{} and \SI{38}{\giga\hertz}. 
    However, this requires the \gls{en} device antenna to have a $(38/3.8)^2\!=\!\SI{20}{\dB}$ larger directivity than the \SI{3.8}{\giga\hertz} device, which in turn
    requires some form of physical or electronic beam steering.
\addtocounter{footnote}{1}
    \item In the physically small \SI{38}{\giga\hertz} system, the maximum power density lies close to the transmitting array. 
    In fact it is also much larger than in the \SI{3.8}{\giga\hertz} case. 
    At \SI{3.8}{\giga\hertz}, the maximum power density is shifted towards the EN device, enabled through the range-dependent array gain pattern\footnotemark[3]\!\cite{Bjornson21Asilomar} of the physically large aperture in the near-field. 
\end{itemize}
Both these observations are consequences of the physically large aperture of the \SI{3.8}{\giga\hertz} system.
While both systems have the same electrical aperture --- which is defined as the physical aperture area normalized by the squared wavelength --- the physical aperture of the \SI{3.8}{\giga\hertz} system is $38/3.8\!=\!10$ times larger in each dimension than that of the \SI{38}{\giga\hertz} system. 
Note that the angular beamwidth is the same in both systems; this is because their electrical aperture in wavelengths squared (and number of antennas) is the same. 

\footnotetext[3]{Demonstrated later using real-world measurements in Fig.\,\ref{fig:GainPattern}\,b).}

Qualitatively, the power densities will look the same at \SI{3.8}{\giga\hertz} and \SI{38}{\giga\hertz}, \emph{if the geometry is scaled correspondingly}. This is a consequence of the scaling invariance of the wave equation. 
Specifically, suppose we move the \gls{en} device ten times closer to the array (\SI{1.23}{\metre} away) and scale the dimensions in Fig.\,\ref{fig:BFsim}\,a)  by $1/10$. 
We then obtain \emph{qualitatively} the same behavior as in Fig.\,\ref{fig:BFsim}\,b): the maximum power density occurs near to the device. 
Also, the power density is \SI{20}{\dB} higher in this case than in Fig.\,\ref{fig:BFsim}\,b), which leads to the same harvested power, even without the need for a directional antenna at the device. But the ultimate consequence is a reduction of the achieved range by a factor of ten, which makes the system impractical.

The presence of multipath propagation effectively increases the physical aperture:
the multipath propagation at specular surfaces can be modeled as virtual mirror arrays (see Fig.\,\ref{fig:MagazineScenario}). 
To illustrate this phenomenon, we consider the \SI{3.8}{\giga\hertz} system operating in the hallway scenario schematically depicted in Fig.\,\ref{fig:MagazineScenario}. 
Figures~\ref{fig:BFsim}\,d) and~\ref{fig:BFsim}\,e) show the power density separately for two different multipath components, and Fig.\,\ref{fig:BFsim}\,f) shows the power density resulting from all multipath components and the \gls{los} path combined.
We observe the following:
\begin{itemize}
   \item The radiated waves from all antennas, after undergoing multipath propagation,
   combine constructively at the \gls{en} device.  This is accomplished by the conjugate beamforming, which automatically accounts for multipath propagation and near-field effects.

   \item The presence of multipath effectively enlarges the physical aperture, shifting the
   maximum power density to the close vicinity of the \gls{en} device in Fig.\,\ref{fig:BFsim}\,f).
\end{itemize}
In summary, a physically large aperture leads to a \emph{decreasing} power density near the array and a maximum power density close to the device.
This should be contrasted with physically small apertures, which exhibit an \emph{increasing} power density near the array. 

The choice of operating frequency is fundamentally tied to the realizable physical aperture size for both the infrastructure and the device.
To maintain a constant physical aperture while increasing the frequency, the number of antennas in the array must be increased.
Although this approach can potentially improve both the \gls{wpt} efficiency and the achievable range, there are fundamental limitations to such frequency scaling:
(i) The focal region’s effective width narrows, which eventually makes the beam-acquisition procedure prohibitively difficult. 
(ii) A constant-aperture \emph{receive} antenna becomes more directive at higher frequencies, restricting efficient reception to specific directions, or requiring beam steering, which is considered impractical at an EN device. 
All of these factors must be carefully considered when designing efficient and scalable \gls{wpt} systems.

\ifthenelse{\equal{\IEEEversion}{true}}
{
Supported by the real-world synthetic aperture measurements in Fig.\,\ref{fig:MagazineScenario}, we validate the advantages of physically large apertures for \gls{wpt}, leading to the following contributions:
First, we investigate regulatory limits and show how physically large apertures can aid regulatory compliance. 
If the size of the aperture is sufficiently large, the global maximum of power density can be shifted to the location of the \gls{en} device. 
We derive the achievable power budgets w.r.t. human exposure regulations and illustrate that operating at sub-10\,GHz frequencies can increase the receivable power from the microwatt to the milliwatt range.
We show that this feature of physically large apertures is enabled by the range-dependent near-field gain pattern.
We demonstrate that power-optimal beam focusing naturally exploits multipath propagation.
This effectively enlarges the transmit aperture~\cite{PizzoTWC2023}, improving both the \gls{wpt} efficiency and regulatory compliance. 
The unprecedented receive power levels will necessitate new \acrlong{ic} architectures that enable \gls{en} devices to operate efficiently over a wide dynamic range.
We further address two fundamental challenges of batteryless \gls{en} device operation: The initial access problem and the problem of self-interference mitigation.
}
{
Supported by the real-world synthetic aperture measurements in Fig.\,\ref{fig:MagazineScenario}, we validate the advantages of physically large apertures for \gls{wpt}, leading to the following contributions:
In~\cref{sec:power-budgets}, we investigate regulatory limits and show how physically large apertures can aid regulatory compliance. 
If the size of the aperture is sufficiently large, the global maximum of power density can be shifted to the location of the \gls{en} device. 
We derive the achievable power budgets w.r.t. human exposure regulations and illustrate that operating at sub-10\,GHz frequencies can increase the receivable power from the microwatt to the milliwatt range.
In~\cref{sec:compliance}, we show that this feature of physically large apertures is enabled by the range-dependent near-field gain pattern.
In~\cref{sec:NF-WPT-Multipath}, we demonstrate that power-optimal beam focusing naturally exploits multipath propagation.
This effectively enlarges the transmit aperture~\cite{PizzoTWC2023}, improving both the \gls{wpt} efficiency and regulatory compliance. 
In~\cref{sec:wpt-service}, we show that unprecedented receive power levels will necessitate new \acrlong{ic} architectures that enable \gls{en} devices to operate efficiently over a wide dynamic range.
In Section\,\ref{sec:operation-challenges}, we address two fundamental challenges of batteryless \gls{en} device operation: The initial access problem and the problem of self-interference mitigation.
}

\section{The Potential of Physical Aperture}
Being large w.r.t. the propagation distances of interest, physically large apertures typically operate in the array near-field.
Unlike the range-independent far-field (plane-wave) array gain pattern, the near-field (spherical-wave) array gain pattern becomes range-dependent~\cite[p.\,25\,f.]{D4_1}.
In a communication context, this range-dependence enables spatial separation of users~\cite{Ramenzani23NF,Zhang23NF-Magazine}. 
However, we show that it also significantly aids \gls{wpt} with physically large apertures through reduced interference, better regulatory compliance, and higher power budgets.

\subsection{Power Density Regulations and Power Budgets}\label{sec:power-budgets}
Regulations typically limit two quantities:
maximum power density and \gls{eirp}, that is, the product of antenna gain and transmit power.
The European Council Recommendation 1999/519/EC limits the maximum power density allowed in the European Union, while the Federal Communications Commission (FCC) 47 CFR §1.1310 limits the maximum power density allowed in the United States.
The \emph{reference levels} in the former, and \emph{maximum permissible exposure} in the latter, limit the power density at a maximum of $S_\mrm{max} = \SI{10}{\watt\per\square\metre}$ for frequencies higher than \SI{2}{\giga\hertz}, and \SI{1.5}{\giga\hertz}, respectively.
The limitation of the maximum power density is motivated by human exposure safety. 
It is a useful quantity for regulating near-field beam focusing, which can be easily evaluated spatially, and has therefore received attention in related work on \gls{wpt} in distributed radio infrastructures~\cite{Lopez22RadioStripesWPT}.

Considering only the device side and assuming a power density limit of $S_\mrm{max} = \SI{10}{\watt\per\square\metre}$, the maximum receivable power would depend only on the effective aperture $A_\mrm{r}$ of the receive antenna. 
Under this assumption, an \gls{en} device equipped with an isotropic receive antenna could optimally attain the maximum receivable powers $P_\mrm{r,max}$ listed in Table\,\ref{tab:max-receivable-powers}.
As the table shows, \gls{wpt} at sub-10\,GHz frequencies can increase the maximum regulatory-compliant receive power from what was conventionally located in the microwatt range~\cite{Hamed23IoT,Zhang23NF-Magazine} to the milliwatt range. 

\begin{table}
			\centering
			\caption
			[Maximum regulatory compliant receivable power.]
			{Maximum power receivable through an isotropic receive antenna at an incident power density of \SI{10}{\watt\per\metre\squared} for a range of different frequencies. 
                }%
\label{tab:max-receivable-powers}
			\begin{tabularx}{0.54\columnwidth}{@{}c|ccc}
    			\toprule
    			\multirow{2}{5.5em}{\,\textbf{Frequency}} & \multicolumn{2}{c}{\textbf{Isotropic antenna}} 
                    \\
		 	   & $A_\mrm{r}$ &  $P_\mrm{r,max}$ 
                    \\
   			\midrule
    			\SI{917}{\mega\hertz} & 
    			    \SI{85}{\centi\metre\squared} &
    			    \SI{85}{\milli\watt}\textsuperscript{*} 
                        \\
                \SI{2.4}{\giga\hertz} & 
    			    \SI{12}{\centi\metre\squared} &
    			    \SI{12}{\milli\watt} 
                        \\
                \SI{3.8}{\giga\hertz} & 
    			    \SI{5}{\centi\metre\squared} &
    			    \SI{5}{\milli\watt} 
                        \\
                \SI{6.0}{\giga\hertz} & 
    			    \SI{2}{\centi\metre\squared} &
    			    \SI{2}{\milli\watt}
                \\
                \SI{30}{\giga\hertz} & 
    			    \SI{0.08}{\centi\metre\squared} &
    			    \SI{0.08}{\milli\watt} 
                        \\
    			\bottomrule
			\end{tabularx}\\
   {\centering
    \textsuperscript{*}\scriptsize In 1999/519/EC,
        $S_\mrm{max}$ decreases linearly with $f$ below \SI{2}{\giga\hertz}. \\
        In FCC 47 CFR §1.1310, $S_\mrm{max}$ decreases linearly with $f$ below \SI{1.5}{\giga\hertz}.
    }
\end{table}

We evaluate the regulatory compliance of a single array in the hallway scenario depicted in Fig.\,\ref{fig:MagazineScenario}. 
In this scenario, we measured CSI --- defined as the vector of transmission coefficients (S‑parameters) from the $(40\times25)$ transmit antennas to the receive antenna --- via a \gls{vna} (see~\cite[Sec.\,V]{Deutschmann23ICC},\cite[Sec.\,III]{wilding2023propagation}), and modeled (i.e., predicted) it through an image source model~\cite{PizzoTWC2023}, where mirror sources are used to represent first-order \glspl{smc}, that is, specular reflections with equal incidence and reflection angles.
Being our best estimate, we henceforth treat measured \gls{csi} as \emph{perfect} \gls{csi}, whereas the geometrically modeled \gls{csi} --- based on a spherical-wavefront (near-field) \gls{smc} channel model~\cite[Sec.\,IV]{Deutschmann23ICC} --- is denoted as \emph{predicted} \gls{csi}.

Physically large apertures aid compliance with power density limits, as near-field beam focusing positively affects the spatial distribution of the power density, particularly in the proximity of the infrastructure.
Fig.\,\ref{fig:BFsim} shows the impact of beamforming via individual \glspl{smc} $k \in \{1 \, \dots \, 4\}$:

Using only the \gls{los} (i.e., $k\!=\!1$), a strong beam is directed towards the \gls{en} device (see Fig.\,\ref{fig:BFsim}\,b). 
The global maximum power density is located at some distance before the device, while the power density decreases strongly towards the array.   
Our measurements indicate that efficient \gls{wpt} can also be performed by exploiting \glspl{smc}, enabling \gls{nlos} beam focusing which can bypass \gls{olos} conditions.
For instance, using component $k\!=\!3$, that is the reflection caused by the wall in the negative $y$-direction,
efficient transmission remains achievable, despite lower antenna gains, longer propagation distances, and wall-induced attenuation (see Fig.\,\ref{fig:BFsim}\,d).
The latter two are losses that \glspl{smc} typically incur according to electromagnetic theory~\cite{PizzoTWC2023}. 
Beamforming via component $k\!=\!2$ is inefficient because the corresponding incidence angle --- with vertical polarization --- is near the Brewster angle, where the reflection coefficient at the floor is minimal (see Fig.\,\ref{fig:BFsim}\,c).
\emph{Visibility} is a characteristic of physically large apertures that appears in combination with obstructions or the limited extent of reflective surfaces that produce \glspl{smc}~\cite{wilding2023propagation}. 
Due to the limited extent of the respective wall, only a portion of the mirror source $k\!=\!4$ (in the positive $y$-direction) is visible from the perspective of the \gls{en} device. 
As a consequence, its \gls{wpt} efficiency and aperture are effectively reduced, resulting in a wider beam (see Fig.\,\ref{fig:BFsim}\,e).
Jointly using all \glspl{smc} $k\in\{1\,\dots\,4\}$ in a multibeam transmission (see Fig.\,\ref{fig:BFsim}\,f)
leverages multipath propagation, effectively increasing the physical size of the transmit aperture, which results in:
\begin{itemize} 
    \item A narrower focal region and a shift of the global maximum of power density to the location of the \gls{en} device
    \item Higher \gls{wpt} efficiency
    \item Spatially lower power density levels outside the focal region, promoting better regulatory compliance
\end{itemize}

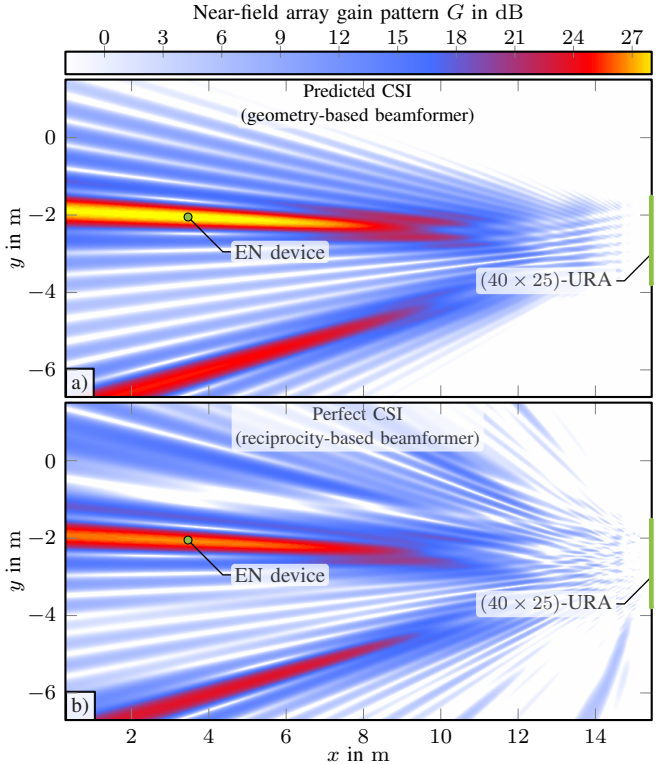
\begin{figure}
        \def\datapath{./figures}
        \setlength{\figurewidth}{1.6\columnwidth}
        \hspace{-1mm}
%
%
\definecolor{mycolor1}{rgb}{0.55294,0.75294,0.27059}%

\pgfplotsset{every axis/.append style={
  label style={font=\footnotesize},
  legend style={font=\footnotesize},
  tick label style={font=\footnotesize},
}}

\begin{tikzpicture}

\begin{axis}[%
width=0.55\figurewidth,
height=0.298\figurewidth,
at={(0\figurewidth,0\figurewidth)},
scale only axis,
point meta min=-2,
point meta max=28,
axis line style = thick,	
line cap = round,
line join = round,
axis on top,
xmin=0.29,
xmax=15.4559999999999,
xticklabel=\empty,                      
ymin=-6.7,
ymax=1.5,
ylabel={$y$ in \SI{}{\metre}},        
ylabel style={yshift=-2mm},        
axis background/.style={fill=white},
colormap={mymap}{[1pt] rgb(0pt)=(0.996221,0.996221,0.996221); rgb(36pt)=(0.57167,0.663465,0.996221); rgb(37pt)=(0.559877,0.654221,0.996221); rgb(38pt)=(0.548083,0.644978,0.996221); rgb(39pt)=(0.53629,0.635735,0.996221); rgb(40pt)=(0.524497,0.626492,0.996221); rgb(41pt)=(0.512704,0.617248,0.996221); rgb(42pt)=(0.500911,0.608005,0.996221); rgb(43pt)=(0.489118,0.598762,0.996221); rgb(44pt)=(0.477325,0.589519,0.996221); rgb(45pt)=(0.465532,0.580275,0.996221); rgb(46pt)=(0.453739,0.571032,0.996221); rgb(47pt)=(0.441945,0.561789,0.996221); rgb(48pt)=(0.430152,0.552546,0.996222); rgb(49pt)=(0.418359,0.543302,0.996221); rgb(50pt)=(0.406567,0.534059,0.996221); rgb(51pt)=(0.394771,0.524816,0.996223); rgb(52pt)=(0.382981,0.515573,0.99622); rgb(53pt)=(0.371194,0.506328,0.996215); rgb(54pt)=(0.359374,0.49709,0.99624); rgb(55pt)=(0.347601,0.487843,0.996221); rgb(56pt)=(0.335907,0.478582,0.996126); rgb(57pt)=(0.323811,0.469392,0.996416); rgb(58pt)=(0.312052,0.460143,0.996384); rgb(59pt)=(0.301781,0.450632,0.994923); rgb(60pt)=(0.286814,0.441948,0.997969); rgb(61pt)=(0.272906,0.433077,1); rgb(62pt)=(0.287748,0.419142,0.974432); rgb(63pt)=(0.320483,0.402054,0.931687); rgb(64pt)=(0.347665,0.385945,0.894273); rgb(65pt)=(0.37359,0.370058,0.858067); rgb(66pt)=(0.401016,0.353906,0.820418); rgb(67pt)=(0.428144,0.337806,0.783056); rgb(68pt)=(0.455101,0.321737,0.745858); rgb(69pt)=(0.482176,0.305647,0.708547); rgb(70pt)=(0.509242,0.289558,0.671245); rgb(71pt)=(0.536289,0.273473,0.63396); rgb(72pt)=(0.563344,0.257386,0.596668); rgb(73pt)=(0.5904,0.241299,0.559375); rgb(74pt)=(0.617455,0.225213,0.522084); rgb(75pt)=(0.644509,0.209126,0.484792); rgb(76pt)=(0.671564,0.19304,0.447501); rgb(77pt)=(0.698619,0.176953,0.410209); rgb(78pt)=(0.725675,0.160861,0.372915); rgb(79pt)=(0.752726,0.14479,0.335629); rgb(80pt)=(0.779781,0.128702,0.298337); rgb(81pt)=(0.806856,0.112535,0.261017); rgb(82pt)=(0.833868,0.0966159,0.223784); rgb(83pt)=(0.860895,0.080641,0.186531); rgb(84pt)=(0.88823,0.0634504,0.148854); rgb(85pt)=(0.9149,0.0488806,0.112092); rgb(86pt)=(0.941208,0.0357377,0.0758298); rgb(87pt)=(0.971794,0.00572876,0.03367); rgb(88pt)=(0.996221,0,0); rgb(89pt)=(0.998849,0.0802077,0.0036217); rgb(90pt)=(0.995317,0.184696,0.00124634); rgb(91pt)=(0.996064,0.272317,0.000217044); rgb(92pt)=(0.996449,0.361366,0.000313345); rgb(93pt)=(0.996169,0.453036,7.27391e-05); rgb(94pt)=(0.996197,0.54349,3.37298e-05); rgb(95pt)=(0.99624,0.633886,2.54196e-05); rgb(96pt)=(0.996219,0.724534,3.42187e-06); rgb(97pt)=(0.996217,0.815106,5.47498e-06); rgb(98pt)=(0.996226,0.905636,6.84373e-06); rgb(99pt)=(0.996221,0.996221,0)},
colorbar horizontal,
colorbar style={%
at={(0,1.088)}, %
anchor=north west, %
height=0.3cm,%
xtick pos=upper, %
xticklabel pos=upper,%
xticklabel style={yshift = -0.1cm},%
xtick distance={3}, %
tick align=center, %
axis line style = thick,	
yshift = -0.12cm, %
xlabel={Near-field array gain pattern $G$ in \SI{}{\dB}},%
line cap = round,
line join = round,
xlabel style={yshift=-1.5mm}}
]
\addplot [forget plot] graphics [xmin=0.28, xmax=15.4, ymin=-6.71, ymax=1.51] {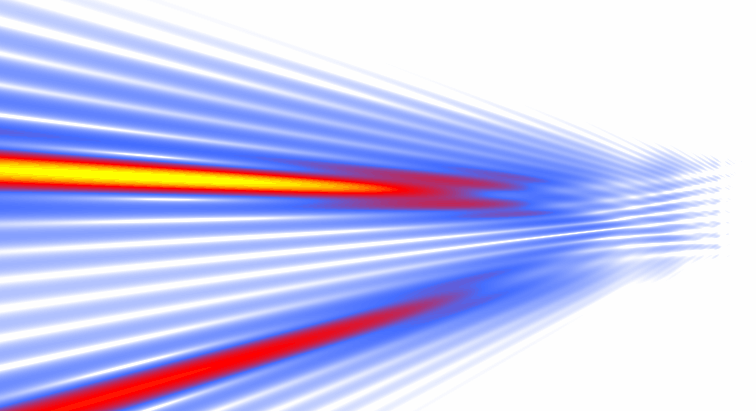};
\addplot [color=mycolor1, only marks, mark size=0.7pt, mark=*, mark options={solid, mycolor1}, forget plot]
  table[row sep=crcr]{%
15.446	-1.53580769544671\\
15.446	-1.59320217260329\\
15.446	-1.65059664975987\\
15.446	-1.70799112691645\\
15.446	-1.76538560407303\\
15.446	-1.82278008122961\\
15.446	-1.88017455838618\\
15.446	-1.93756903554276\\
15.446	-1.99496351269934\\
15.446	-2.05235798985592\\
15.446	-2.1097524670125\\
15.446	-2.16714694416908\\
15.446	-2.22454142132566\\
15.446	-2.28193589848224\\
15.446	-2.33933037563882\\
15.446	-2.3967248527954\\
15.446	-2.45411932995197\\
15.446	-2.51151380710855\\
15.446	-2.56890828426513\\
15.446	-2.62630276142171\\
15.446	-2.68369723857829\\
15.446	-2.74109171573487\\
15.446	-2.79848619289145\\
15.446	-2.85588067004803\\
15.446	-2.91327514720461\\
15.446	-2.97066962436118\\
15.446	-3.02806410151776\\
15.446	-3.08545857867434\\
15.446	-3.14285305583092\\
15.446	-3.2002475329875\\
15.446	-3.25764201014408\\
15.446	-3.31503648730066\\
15.446	-3.37243096445724\\
15.446	-3.42982544161382\\
15.446	-3.4872199187704\\
15.446	-3.54461439592697\\
15.446	-3.60200887308355\\
15.446	-3.65940335024013\\
15.446	-3.71679782739671\\
15.446	-3.77419230455329\\
15.446	-3.77419230455329\\
};


\draw[color=black, line width=0.5pt]    
(15.45,-3.000) -- %
(14.75,-3.700) -- %
(14.55,-3.700); %
\node[left, align=right ,font={\footnotesize},fill=white,
opacity=0.75,inner sep=0.5mm, xshift=0.0mm, rounded corners=0.5mm] at %
(axis cs:14.55,-3.700){\scalebox{0.9}{$(40\times 25)$}-URA};

\draw[color=black, line width=0.5pt]    
(3.462,-2.050) -- %
(4.362,-2.950) -- %
(4.562,-2.950); %
\node[right, align=left ,font={\footnotesize},fill=white,
opacity=0.75,inner sep=0.5mm, xshift=0.0mm, rounded corners=0.5mm] at %
(axis cs:4.562,-2.950){EN device};

\addplot[only marks, mark=*, mark options={}, mark size=1.5000pt, color=black, fill=RDlightgreen, forget plot] table[row sep=crcr]{%
x	y\\
3.46166087184211	-2.05024565388158\\
}; 

\node[below, align=center ,font={\footnotesize},fill=white,
opacity=0.5,text opacity = 1,inner sep=0.5mm, xshift=0.0mm, rounded corners=0.5mm] at %
(axis cs:7.85,1.5){\scalebox{0.9}{Predicted CSI}\\\scalebox{0.9}{(geometry-based beamformer)}};  

\node[right, align=left,font={\footnotesize},fill=white,
opacity=0.9,inner sep=0.75mm, xshift=0.5mm, draw, line width = 0.8pt] at %
(axis cs:0.18,-6.35){a)};

\end{axis}
\end{tikzpicture}%
        \vspace{-1.8mm}%
        \hspace{-1mm}
%
%
\definecolor{mycolor1}{rgb}{0.55294,0.75294,0.27059}%

\pgfplotsset{every axis/.append style={
  label style={font=\footnotesize},
  legend style={font=\footnotesize},
  tick label style={font=\footnotesize},
}}

\begin{tikzpicture}

\begin{axis}[%
width=0.55\figurewidth,
height=0.298\figurewidth,
at={(0\figurewidth,0\figurewidth)},
scale only axis,
point meta min=-2,
point meta max=28,
axis line style = thick,	
axis on top,
line cap = round,
line join = round,
xmin=0.29,
xmax=15.4559999999999,
xlabel={$x$ in \SI{}{\metre}},        
xlabel style={yshift=0.2cm},
ymin=-6.7,
ymax=1.5,
ylabel={$y$ in \SI{}{\metre}},        
ylabel style={yshift=-2mm},        
axis background/.style={fill=white},
colormap={mymap}{[1pt] rgb(0pt)=(0.996221,0.996221,0.996221); rgb(36pt)=(0.57167,0.663465,0.996221); rgb(37pt)=(0.559877,0.654221,0.996221); rgb(38pt)=(0.548083,0.644978,0.996221); rgb(39pt)=(0.53629,0.635735,0.996221); rgb(40pt)=(0.524497,0.626492,0.996221); rgb(41pt)=(0.512704,0.617248,0.996221); rgb(42pt)=(0.500911,0.608005,0.996221); rgb(43pt)=(0.489118,0.598762,0.996221); rgb(44pt)=(0.477325,0.589519,0.996221); rgb(45pt)=(0.465532,0.580275,0.996221); rgb(46pt)=(0.453739,0.571032,0.996221); rgb(47pt)=(0.441945,0.561789,0.996221); rgb(48pt)=(0.430152,0.552546,0.996222); rgb(49pt)=(0.418359,0.543302,0.996221); rgb(50pt)=(0.406567,0.534059,0.996221); rgb(51pt)=(0.394771,0.524816,0.996223); rgb(52pt)=(0.382981,0.515573,0.99622); rgb(53pt)=(0.371194,0.506328,0.996215); rgb(54pt)=(0.359374,0.49709,0.99624); rgb(55pt)=(0.347601,0.487843,0.996221); rgb(56pt)=(0.335907,0.478582,0.996126); rgb(57pt)=(0.323811,0.469392,0.996416); rgb(58pt)=(0.312052,0.460143,0.996384); rgb(59pt)=(0.301781,0.450632,0.994923); rgb(60pt)=(0.286814,0.441948,0.997969); rgb(61pt)=(0.272906,0.433077,1); rgb(62pt)=(0.287748,0.419142,0.974432); rgb(63pt)=(0.320483,0.402054,0.931687); rgb(64pt)=(0.347665,0.385945,0.894273); rgb(65pt)=(0.37359,0.370058,0.858067); rgb(66pt)=(0.401016,0.353906,0.820418); rgb(67pt)=(0.428144,0.337806,0.783056); rgb(68pt)=(0.455101,0.321737,0.745858); rgb(69pt)=(0.482176,0.305647,0.708547); rgb(70pt)=(0.509242,0.289558,0.671245); rgb(71pt)=(0.536289,0.273473,0.63396); rgb(72pt)=(0.563344,0.257386,0.596668); rgb(73pt)=(0.5904,0.241299,0.559375); rgb(74pt)=(0.617455,0.225213,0.522084); rgb(75pt)=(0.644509,0.209126,0.484792); rgb(76pt)=(0.671564,0.19304,0.447501); rgb(77pt)=(0.698619,0.176953,0.410209); rgb(78pt)=(0.725675,0.160861,0.372915); rgb(79pt)=(0.752726,0.14479,0.335629); rgb(80pt)=(0.779781,0.128702,0.298337); rgb(81pt)=(0.806856,0.112535,0.261017); rgb(82pt)=(0.833868,0.0966159,0.223784); rgb(83pt)=(0.860895,0.080641,0.186531); rgb(84pt)=(0.88823,0.0634504,0.148854); rgb(85pt)=(0.9149,0.0488806,0.112092); rgb(86pt)=(0.941208,0.0357377,0.0758298); rgb(87pt)=(0.971794,0.00572876,0.03367); rgb(88pt)=(0.996221,0,0); rgb(89pt)=(0.998849,0.0802077,0.0036217); rgb(90pt)=(0.995317,0.184696,0.00124634); rgb(91pt)=(0.996064,0.272317,0.000217044); rgb(92pt)=(0.996449,0.361366,0.000313345); rgb(93pt)=(0.996169,0.453036,7.27391e-05); rgb(94pt)=(0.996197,0.54349,3.37298e-05); rgb(95pt)=(0.99624,0.633886,2.54196e-05); rgb(96pt)=(0.996219,0.724534,3.42187e-06); rgb(97pt)=(0.996217,0.815106,5.47498e-06); rgb(98pt)=(0.996226,0.905636,6.84373e-06); rgb(99pt)=(0.996221,0.996221,0)},
]
\addplot [forget plot] graphics [xmin=0.28, xmax=15.4, ymin=-6.71, ymax=1.51] {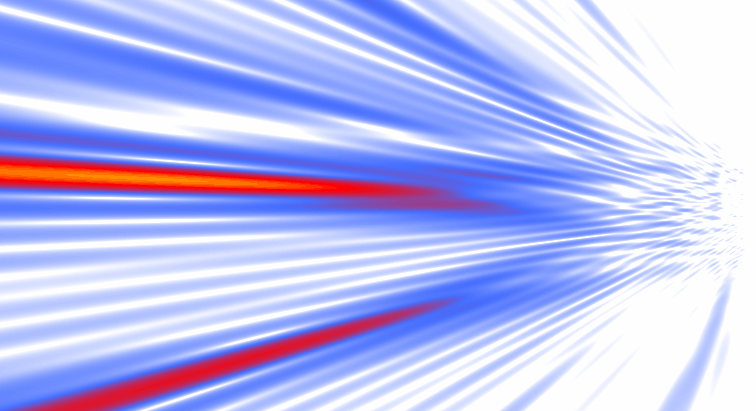};
\addplot [color=mycolor1, only marks, mark size=0.7pt, mark=*, mark options={solid, mycolor1}, forget plot]
  table[row sep=crcr]{%
15.446	-1.53580769544671\\
15.446	-1.59320217260329\\
15.446	-1.65059664975987\\
15.446	-1.70799112691645\\
15.446	-1.76538560407303\\
15.446	-1.82278008122961\\
15.446	-1.88017455838618\\
15.446	-1.9375690355427\\
15.446	-1.99496351269934\\
15.446	-2.05235798985592\\
15.446	-2.1097524670125\\
15.446	-2.16714694416908\\
15.446	-2.22454142132566\\
15.446	-2.28193589848224\\
15.446	-2.33933037563882\\
15.446	-2.3967248527954\\
15.446	-2.45411932995197\\
15.446	-2.51151380710855\\
15.446	-2.56890828426513\\
15.446	-2.62630276142171\\
15.446	-2.68369723857829\\
15.446	-2.74109171573487\\
15.446	-2.79848619289145\\
15.446	-2.85588067004803\\
15.446	-2.91327514720461\\
15.446	-2.97066962436118\\
15.446	-3.02806410151776\\
15.446	-3.08545857867434\\
15.446	-3.14285305583092\\
15.446	-3.2002475329875\\
15.446	-3.25764201014408\\
15.446	-3.31503648730066\\
15.446	-3.37243096445724\\
15.446	-3.42982544161382\\
15.446	-3.4872199187704\\
15.446	-3.54461439592697\\
15.446	-3.60200887308355\\
15.446	-3.65940335024013\\
15.446	-3.71679782739671\\
15.446	-3.77419230455329\\
15.446	-3.77419230455329\\
};


\draw[color=black, line width=0.5pt]    
(15.45,-3.000) -- %
(14.75,-3.700) -- %
(14.55,-3.700); %
\node[left, align=right ,font={\footnotesize},fill=white,
opacity=0.75,inner sep=0.5mm, xshift=0.0mm, rounded corners=0.5mm] at %
(axis cs:14.55,-3.700){\scalebox{0.9}{$(40\times 25)$}-URA};

\draw[color=black, line width=0.5pt]    
(3.462,-2.050) -- %
(4.362,-2.950) -- %
(4.562,-2.950); %
\node[right, align=left ,font={\footnotesize},fill=white,
opacity=0.75,inner sep=0.5mm, xshift=0.0mm, rounded corners=0.5mm] at %
(axis cs:4.562,-2.950){EN device};

\addplot[only marks, mark=*, mark options={}, mark size=1.5000pt, color=black, fill=RDlightgreen, forget plot] table[row sep=crcr]{%
x	y\\
3.46166087184211	-2.05024565388158\\
}; 

\node[below, align=center ,font={\footnotesize},fill=white,
opacity=0.75,inner sep=0.5mm, xshift=0.0mm, rounded corners=0.5mm] at %
(axis cs:7.85,1.5){\scalebox{0.9}{Perfect CSI}\\\scalebox{0.9}{(reciprocity-based beamformer)}};  

\node[right, align=left,font={\footnotesize},fill=white,
opacity=0.9,inner sep=0.75mm, xshift=0.5mm, draw, line width = 0.8pt] at %
(axis cs:0.18,-6.35){b)};

\end{axis}
\end{tikzpicture}%
        \vspace{-2mm}%
        \caption{Range-dependent near-field array gain pattern~\cite[eq.\,(3.8)]{D4_1} computed for geometry-based predicted \gls{csi} (a) and ``perfect'' \gls{csi} (b) from measurements. 
        The geometry-based spherical-wavefront beamformer generates a gain pattern similar to that of perfect \gls{csi} but does not leverage all available multipath components.
        }
        \label{fig:GainPattern}
\end{figure}

If the physical aperture of the transmit array becomes sufficiently large, it is possible to generate the global power density maximum at the position of the \gls{en} device (see \cite{Lopez22RadioStripesWPT}). 
Under this assumption, the power receivable by the \gls{en} device merely depends on the power density regulations and the effective aperture $A_\mrm{r}$ of its receiving antenna. 
Therefore, for physically large apertures, the power budget at the \gls{en} device side becomes practically independent of the distance from the infrastructure, and the power budgets in Table\,\ref{tab:max-receivable-powers} can be obtained.
This represents a clear paradigm shift from conventional \gls{rf} \gls{wpt} systems operating with physically small apertures.  

The given scenario shows the performance peculiarities mentioned above in the example of a single physically large array.
Performing \gls{wpt} at \SI{3.8}{\giga\hertz}, the \gls{en} device could receive up to \SI{10}{\milli\watt} if the array transmits at a total power of \SI{10}{\watt}, given perfect \gls{csi}, while still adhering to the power density limit of $\SI{10}{\watt\per\square\metre}$, see Fig.\,\ref{fig:BFsim}\,f).

\subsection{\gls{eirp} Limits and Near-Field Array Gain Pattern}\label{sec:compliance}
The European Commission Decision 2006/771/EC limits \gls{eirp} in the European Union, while FCC 47 CFR §15.407 limits \gls{eirp} in the United States, the main motivation being the reduction of interference. 
For \gls{siso} systems, the \gls{eirp} limit is the product of antenna gain and transmit power and constitutes a quantity that is practical to evaluate.
While far-field array gain patterns are solely a function of elevation and azimuth angles, near-field beam focusing results in array gain patterns that are range-dependent. 
This aids compliance with gain regulations as the maximum is usually focused at the device location. 
Thus, for near-field beam focusing with \gls{miso} systems, the existing regulations do not reflect how array gain affects the \gls{eirp}.
\setcounter{romanNumerals}{1}
Specifically, 
high near-field array gains at the device position can still result in (\roman{romanNumerals})\refstepcounter{romanNumerals} low far-field array gains~\cite[p.\,26]{D4_1} and (\roman{romanNumerals}) very low near-field array gains close to the array.
This demonstrates that near-field beam focusing reduces interference with receivers at both far and close distances.
Thus, near-field beam focusing possibly necessitates adaptations to the EIRP regulations used to limit interference.

Fig.\,\ref{fig:GainPattern} depicts the resulting near-field array gain patterns~\cite[eq.\,(3.8)]{D4_1} when using predicted \gls{csi} and perfect \gls{csi}.
We use conjugate beamforming~\cite{mMIMObook}, that is, normalized vectors of beamforming weights computed from conjugate \gls{csi}. 
The near-field array gain pattern is the inner product of these beamforming vectors with the steering vectors evaluated in the near-field.
Predicting \gls{csi} with the spherical-wavefront \gls{smc} channel model results in a beamformer that exploits strong first-order \glspl{smc} (see Fig.\,\ref{fig:GainPattern}\,a). 
It predominantly uses the \gls{los} $k\!=\!1$ and component $k\!=\!3$. 
There is barely any power directed to component $k\!=\!4$ due to its limited visibility. 
Any array gain pattern $G$ is upper-bounded by the number of transmit antennas, which is $L\!=\!1000$ in our measurements. 
A reciprocity-based beamformer, given perfect \gls{csi}, results in power-optimal beamforming (see Fig.\,\ref{fig:GainPattern}\,b).
It naturally exploits multipath propagation and makes rich use of the entire channel, including higher-order \glspl{smc} and diffuse reflections. 
In~Fig.\,\ref{fig:GainPattern}\,b), this is visible through some array gain that is diverted in directions other than the two dominant components of~Fig.\,\ref{fig:GainPattern}\,a).
A conventional spherical-wavefront \gls{los} beamformer would generate an array gain of $\SI{30}{\dB}$ at the location of the \gls{en} device,  
while the reciprocity-based beamformer generates a spatial maximum of $G_\mathrm{NF,max}\!=\!\SI{26.3}{\dB}$. 

The far-field array gain pattern reduces to an even lower maximum of $G_\mathrm{FF,max}\!=\!\SI{20.3}{\dB}$, far out~\cite[see Fig.\,5]{Bjornson21Asilomar}, in the direction of the EN device. 
Spherical-wavefront beam focusing results in very low array gains close to the array and decaying array gains ``behind'' the \gls{en} device, which aids regulatory compliance and reduces interference.
These effects become particularly strong with apertures being not only electrically large but also physically large.

\begin{figure*}
\centering
    \setlength{\figurewidth}{0.6\linewidth}
    \setlength{\figureheight}{0.4\linewidth}
    \def\datapath{./figures}
    \input{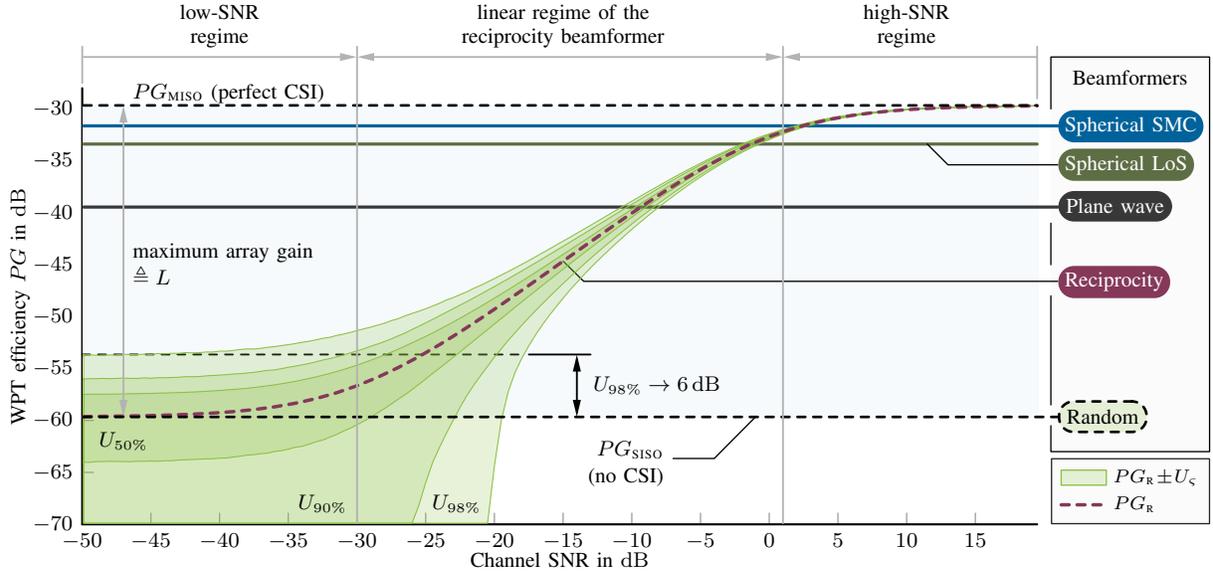}
     \caption{Measured performance comparison of geometry-based beamformers (given the ``true'' device position) with a reciprocity-based beamformer versus the quality of \gls{csi} determined by the channel \gls{snr}~\cite{Deutschmann23ICC}. 
     The efficiency is upper-bounded by $PG_{\textsc{miso}}$ assuming perfect \gls{csi} and leveraging the maximum array gain.
     $PG_{\textsc{siso}}$, the performance baseline, assumes no \gls{csi} and coincides with the expected efficiency of a random beamformer. 
     Accurate geometric near-field channel models increasingly outperform plane-wave or random beamforming using physically large apertures in massive \gls{mimo}.
     }
     \label{fig:BF_comparison} 
 \end{figure*}

\subsection{Near-Field Beam Focusing Strategies}\label{sec:NF-WPT-Multipath}
In practice, perfect \gls{csi} will never be available and deviations from the ``true'' channel will manifest in power losses. 
Taking these \gls{csi} imperfections into account, we discuss the expected efficiency $PG_{\textsc{r}}\!=\!\mathbb{E}\big( PG\big)$ of a reciprocity-based beamformer given noisy \gls{csi}, corrupted by additive spatially white circular Gaussian noise. 
Using the channel \gls{snr} to express the \emph{quality} of \gls{csi}, its efficiency follows three asymptotic regimes (see Fig.\,\ref{fig:BF_comparison}):

In the \emph{low-\gls{snr} regime}, \gls{csi} has a very low quality as it is dominated by noise.
In this regime, depending on the random \gls{csi} realizations, the $PG$ of a reciprocity-based beamformer follows a chi-squared distribution and performs on average no more efficiently than an equivalent \gls{siso} system that uses only a single transmit antenna out of the $L\!=\!1000$ antennas within the \gls{ura}. 
This \gls{siso} \acrlong{pg} ($PG_\textsc{siso}$) may be regarded as a performance baseline. 
However, if all $L$ antennas are used, the power density at the \gls{ura} is reduced for the benefit of regulatory compliance.
The expected efficiency $PG_{\textsc{r}}$ of the distribution is augmented by symmetric intervals $[PG_{\textsc{r}}\!-\!U_\varsigma, PG_{\textsc{r}}\!+\!U_\varsigma]$ in which one random realization of $PG$ is located with probability $\varsigma\!=\!\mathbb{P}\big(PG_{\textsc{r}}\!-U_\varsigma\!\leq\!PG\!\leq\!PG_{\textsc{r}}\!+\!U_\varsigma\big)$.
Fig.~\ref{fig:BF_comparison} shows the intervals computed for probabilities $\varsigma\!\in\!\{50\%,90\%,98\%\}$. 
Within the $98\%$ interval, a gain of up to $\SI{6}{\dB}$ can be attained over $PG_\textsc{siso}$, which corresponds to a \emph{random} beamformer given multiple different random beamforming weight realizations and is achievable regardless of the number of antennas $L$ used. 
In the \emph{linear regime}, the \gls{pg} of a reciprocity-based beamformer is Gaussian distributed and its expected efficiency increases linearly with the \gls{snr} of the \gls{csi}, while the relative \gls{pg} variance decreases. 
Eventually, in the \emph{high-\gls{snr} regime}, the efficiency (i.e., the \gls{pg}) stays Gaussian distributed and saturates at the \gls{miso} \gls{pg}, representing the upper bound on efficiency and leveraging the maximum array gain corresponding to the number of transmit antennas $L$.
This upper bound increases steeply when increasing the number of antennas in massive \gls{mimo} systems, while the baseline performance ($PG_\textsc{siso}$) and the possible performance gain of a random beamformer (\SI{6}{\dB}) do not increase.
Nevertheless, the random beamformer is still a viable option to solve the initial access problem. 

Geometry-based beamforming can leverage a large portion of the available array gain, as indicated by the horizontal lines in Fig.\,\ref{fig:BF_comparison}.
We define three geometry-based channel models of varying accuracy to predict \gls{csi}, assuming a known position of the \gls{en} device~\cite{Deutschmann23ICC}.
The \emph{plane-wave} \gls{los} beamformer corresponds to a conventional far-field beamformer that computes beamforming weights solely as a function of azimuth and elevation angles, which makes it the least complex in terms of model parameters.
Despite outperforming a random beamformer by around \SI{20}{\dB}, it suffers a loss of \SI{10}{\dB} w.r.t. perfect \gls{csi} in the given scenario that cannot be compensated using \glspl{smc}, confirming the necessity of appropriate near-field channel modeling.
The \emph{spherical-wavefront \gls{los}} beamformer (see Fig.\,\ref{fig:BFsim}\,b) suffers a loss of only \SI{4}{\dB} w.r.t. perfect \gls{csi} and represents a good tradeoff between efficiency and model complexity, as it takes the exact distances from each transmit antenna to the device into account. 
The \emph{spherical-wavefront \gls{smc}} beamformer additionally exploits specular reflections to increase its efficiency. 
It outperforms the spherical-wavefront \gls{los} beamformer by \SI{2}{\dB} and achieves a global maximum of power density at the \gls{en} device location, improving its regulatory compliance and maximum receivable power (see Fig.\,\ref{fig:BFsim}\,f). 
Looking at Fig.\,\ref{fig:GainPattern}, the similarity in the array gain patterns shows that the resulting beamformer mimics the reciprocity-based beamformer given perfect \gls{csi} quite successfully. 
We conclude that leveraging multipath propagation is necessary to perform efficient \gls{wpt}, particularly at sub-10\,GHz frequencies. 

Additional closed‑loop beamforming schemes --- including codebook‑based, quantized‑CSI, and iterative feedback approaches --- can also be employed, but all of these are ultimately bounded by the perfect‑CSI efficiency $PG_{\textsc{miso}}$ shown in Fig.\,\ref{fig:BF_comparison}.

\section{Wireless Power Transfer: A Service in 6G}\label{sec:wpt-service}

Some 6G use cases, such as electronic labeling, asset tracking, and real-time inventory, will involve 
 large numbers of distributed \gls{en} devices.
Supporting these use cases requires a radio infrastructure that can deliver \gls{wpt} as a service.

\subsection{Distributed 6G Radio Infrastructures}\label{sec:Infrastructure}

Radio Stripes~\cite{Lopez22RadioStripesWPT} refer to distributed antenna systems integrated into a single cable or stripe, which allow for \gls{cjt} while reducing deployment complexity and cabling costs.
RadioWeaves~\cite{D4_1} take the concept further by embedding radio elements and associated computational resources into everyday surfaces (e.g., walls, ceilings, or even furniture). 
Both are examples of emerging \gls{6g} radio infrastructures with distributed antenna arrays that cooperatively provide hyper-diverse connectivity, computational resources, positioning, and \gls{wpt} to connected devices.
Operating at sub-10\,GHz and involving large numbers of antennas, they inherently form physically large apertures and feature near-field operation.

We have illustrated the fundamental potential of using a single physically large \gls{ura} (Fig.\,1) for \gls{rf} \gls{wpt}.
Building and installing such an array, with $1000$ or more antennas, would be expensive and difficult in practice. 
Furthermore, the power losses incurred in the \gls{rf} chains of a fully digital phased-array implementation may severely decrease the overall system efficiency.
However, we believe that emerging technology, such as Radio Stripes \cite{Lopez22RadioStripesWPT} and RadioWeaves \cite{D4_1},
will eventually be sufficiently cost- and energy-efficient to deploy at this scale.
Particularly, RadioWeaves is a very flexible solution that involves many distributed arrays, 
which together can yield an even larger \emph{combined} array aperture.

\subsection{Energy Neutral Devices}\label{sec:hardware}
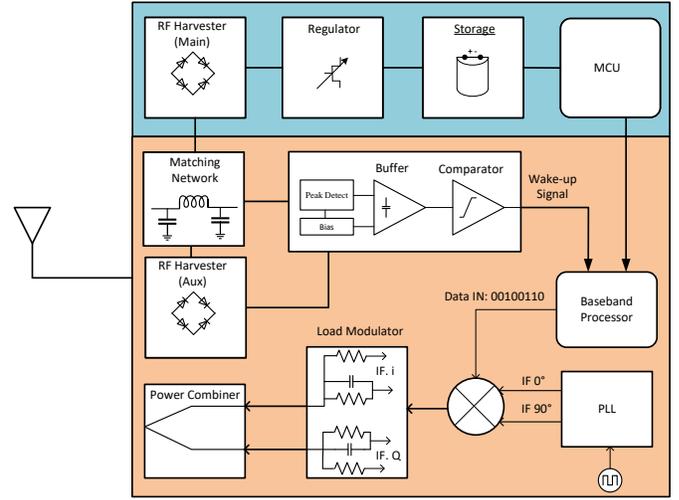
\begin{figure}
        \hspace{-1mm}\begin{tikzpicture}
            \tikzset{
              labelline/.style={
                draw=none,
                fill=none,
                align=center,
                text depth=0pt,
                inner sep=0pt,
                font=\scriptsize
              }
            }
        	\node[draw=none,fill=none] at (0,0){\includegraphics[width=1.0\columnwidth]{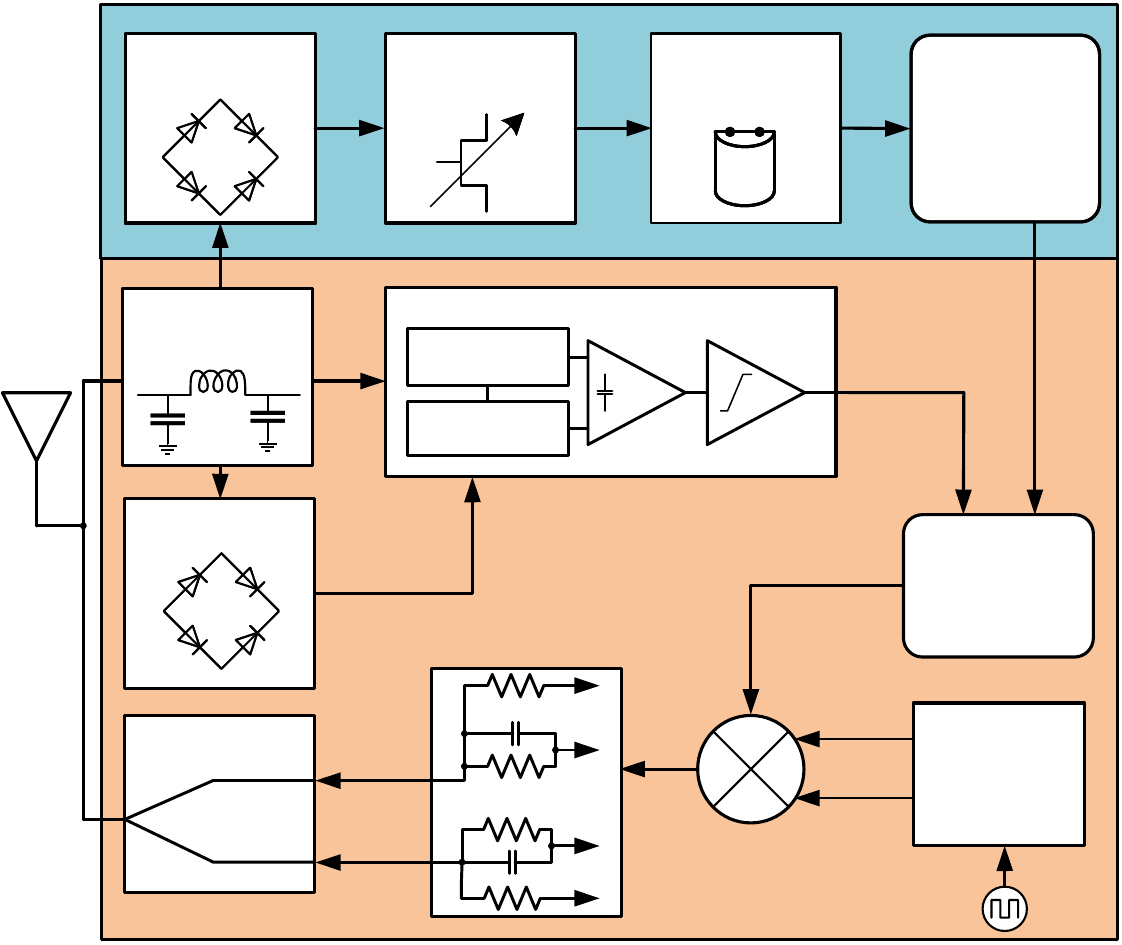}};
        	\node[align=center,draw=none,fill=none] at (-0.56,0.35) {\scriptsize Bias};
        	\node[align=center,draw=none,fill=none] at (-0.56,0.92) {\scriptsize Peak\,Detect};
        	\node[align=center,draw=none,fill=none,anchor=center] at (0.45,1.2) {\scriptsize Buffer};
        	\node[align=center,draw=none,fill=none,anchor=center] at (1.5,1.2) {\scriptsize Comparator};
        	\node[align=center,draw=none,fill=none] at (2.75,0.9) {\scriptsize Wake-up\\[-5pt] \scriptsize Signal};
        	\node[align=center,draw=none,fill=none] at (3.45,-0.85) {\scriptsize Baseband\\[-5pt] \scriptsize Processor};
        	\node[align=center,draw=none,fill=none] at (3.45,-2.4) {\scriptsize Phase-\\[-5pt] \scriptsize Locked\\[-5pt] \scriptsize Loop};
        	\node[align=center,draw=none,fill=none] at (2.42,-1.95) {\scriptsize IF \SI{0}{\degree}};
        	\node[align=center,draw=none,fill=none] at (2.42,-2.41) {\scriptsize IF \SI{90}{\degree}};
        	\node[align=center,draw=none,fill=none] at (2.1,-0.75) {\scriptsize TX Data};
        	\node[align=center,draw=none,fill=none] at (-0.28,-1.4) {\scriptsize Load Modulator (LM)};
        	\node[align=center,draw=none,fill=none] at (0.2,-1.9) {\scriptsize IF I};
        	\node[align=center,draw=none,fill=none] at (0.2,-3.2) {\scriptsize IF Q};
        	\node[align=center,draw=none,fill=none] at (-2.7,-2.15) {\scriptsize Power \\[-6pt] \scriptsize Combiner};
        	\node[align=center,draw=none,fill=none] at (-2.7,-0.35) {\scriptsize ARFH};
        	\node[align=center,draw=none,fill=none] at (-2.7,1.15) {\scriptsize Matching\\[-5pt] \scriptsize Network};
        	\node[align=center,draw=none,fill=none] at (-2.7,3.2) {\scriptsize MRFH};
                    \node[align=center,draw=none,fill=none] at (-1.65,2.95) {\scriptsize DC};
                    \node[align=center,draw=none,fill=none,rotate=90] at (0.4,2.63) {\tiny stabilized ~\tiny voltage};
                    \node[align=center,draw=none,fill=none] at (2.485,2.95) {\scriptsize VCC};
                    \node[align=center,draw=none,fill=none,rotate=90] at (3.6,0.9) {\scriptsize Data Bus};
                    \node[align=center,draw=none,fill=none] at (0.85,-2.05) {\scriptsize IF\\[-5pt] \scriptsize I/Q };
                    \node[align=center,draw=none,fill=none] at (-1.25,-0.82) {\scriptsize DC Bias };
                    \node[align=center,draw=none,fill=none,rotate=90] at (-1.7,0.7) {\scriptsize Demod. Signal };
                    \node[align=center,draw=none,fill=none] at (-1.35,-2.3) {\scriptsize LM I };
                    \node[align=center,draw=none,fill=none] at (-1.35,-2.95) {\scriptsize LM Q };
        	\node[align=center,draw=none,fill=none] at (-0.6,3.2) {\scriptsize Regulator};
        	\node[align=center,draw=none,fill=none] at (1.475,3.1) {\scriptsize Energy\\[-5pt] \scriptsize Storage};
        	\node[align=center,draw=none,fill=none] at (3.51,2.7) {\scriptsize Micro-\\[-5pt] \scriptsize controller\\[-5pt] \scriptsize Unit};
        		\node[align=center,draw=none,fill=none,rotate=90] at (-4.16,1.25) {\scriptsize Antenna};
            \node[left,align=right,draw=none,fill=none] at (3.36,-3.42) {\scriptsize Oscillator};
        \end{tikzpicture}
        \caption{A novel \gls{en} device architecture: 
        Unprecedented power budgets at the device side demand a novel front-end design, with one integrated branch providing high sensitivity during the initial access phase and another off-chip branch providing optimum efficiencies at high input powers.} 
        \label{fig:EMD_arch}
\end{figure}

We showed that \gls{wpt} with physically large apertures lifts receive power budgets from microwatts into the milliwatt range, power levels that prohibit a conventional design with the \gls{rf} harvester integrated on a single silicon die.
A state-of-the-art strategy involves designing \gls{en} devices to operate with maximum efficiency at their device sensitivity, which is the minimum input power required for wake-up, thus maximizing their initial access distance. 
Traditionally, the power harvesting efficiency of the front-end degrades at higher input powers, posing a challenge for simultaneous efficient operation at both the device sensitivity and the maximum power budget.
We propose a front-end design with two branches (see Fig.\,\ref{fig:EMD_arch}): 
The lower branch, termed \gls{arfh}, is dedicated to solving the initial access problem, while the upper branch, termed \gls{mrfh}, is responsible for harvesting the maximum power from the focal region.
The \gls{arfh} must operate at very low sensitivities (e.g., as low as \SI{-25}{\dBm}, typical for \acrshort{rfid} tags) with high conversion efficiency to provide sufficient energy for driving the wake-up signal processing, the baseband processor, and the modulator.
Conversely, the \gls{mrfh} must operate at high voltages (above \SI{5}{\volt}) to deliver power levels up to \SI{100}{\milli\watt} to the storage device to efficiently drive power-consuming functions on the \gls{mcu}.
Consequently, the \gls{arfh} is composed of stages of concatenated charge pumps~\cite{Zoescher17RFID}, and the \gls{mrfh} is fabricated as a single bridge rectifier using Schottky diodes to provide minimal forward voltage drop. 
These contradicting requirements prohibit the integration of both types of harvesters on a single die in a cost-effective way. 
The main contributing factors are \gls{esd} protection, maximum tolerable semiconductor process voltage (usually below \SI{5}{\volt}), and utilization of specific transistor types w.r.t. efficiency versus operational voltage tolerance. 
Hence, the low-power branch is built as an integrated solution, while the high-power branch is located off-chip. Both branches are connected through a dedicated matching network and \gls{esd} protection to the antenna.   

\section{Challenges of Operating Energy-Neutral Devices}\label{sec:operation-challenges}
While batteryless \gls{en} devices constitute a key enabling technology for massive \gls{iot} deployments, the advantages come with several challenges:
(i) reliable initial access before \gls{csi} is available, and (ii) suppression of \gls{dli} when distributed infrastructures (e.g., RadioWeaves or Radio Stripes) are used for backscatter communication and \gls{csi} estimation.

\subsection{Initial Access Strategies}\label{sec:initial-access}
When the infrastructure is put into operation for the first time, it needs to acquire \gls{csi} for downlink beamforming.
As measured \gls{csi} is unavailable before the initial wake-up of an \gls{en} device, our geometry-based beamformers provide an attractive solution for the problem. 
To compensate for a possibly unknown position, a \emph{beam sweep} can be conducted, that is, a space of possible positions can be searched iteratively until the initial wake-up of the \gls{en} device.
Other approaches are codebook-based beamforming, or random beamforming. 
Both belong to the class of \emph{\gls{csi}-free} methods, which have been found beneficial for simultaneously powering a large number of distributed devices~\cite{Lopez19CSIfree}.
The former can achieve good performance if iterated through an exhaustive codebook, but may become prohibitive in massive \gls{mimo} applications due to the large codebook sizes involved.
Because the codebook is a finite, pre‑quantized set that must also cover range‑dependent near‑field beams, its size grows rapidly in the near-field.
Once the devices become active, a pilot can be sent to enable positioning and reciprocity-based beamforming.

\subsection{Direct Link Interference Suppression in Distributed Infrastructures}\label{sec:distributed-infrastrucutes}
A distributed radio infrastructure will involve multiple arrays operating as a jointly coherent, physically large synthetic aperture.
Distributed arrays allow \gls{bibc}, where each of the spatially separated arrays works either as carrier emitter or reader in half-duplex mode. 
This makes the design much less complex than in a monostatic system.

However, \gls{bibc} suffers from \gls{dli} between the carrier emitter and the reader.  
In addition, the power received from the \gls{en} device is much weaker than the \gls{dli}, owing to the double path loss effect on the cascade backscatter channel. 
In the presence of significant \gls{dli}, high-dynamic-range \acrlong{adcs} are required in the reader circuitry to detect the \gls{en} device signal, which are power-hungry devices. 
To enable the use of lower-resolution \acrlong{adcs}, the \gls{dli}  between the carrier emitter and the reader arrays can be suppressed
using sophisticated beamforming and signal processing techniques~\cite{kaplan2023direct}.

\section{Conclusions and Outlook}
We summarize the key contributions of this article:
\begin{itemize} 
    \item \textit{Regulatory compliance}: We have demonstrated how the physical aperture is a fundamental resource for achieving efficient and regulatory-compliant \gls{wpt} in 6G wireless networks.
    Physically large apertures provide high \gls{wpt} efficiency in the near-field focal region while maintaining low power densities outside, even in proximity to the infrastructure, reducing human exposure and improving regulatory compliance.
    \item \textit{Power Budgets}:  Physically large apertures elevate power budgets into the milliwatt range when operating at sub-10\,GHz frequencies, paving the way for a new generation of highly capable \gls{en} devices that can be deployed sustainably at a massive scale. 
    \item \textit{Multipath Propagation}: Conjugate beamforming naturally exploits multipath propagation to increase the effective physical aperture size, which improves the \gls{wpt} efficiency, narrows the focal region, and mitigates interference. 
\end{itemize}
While perfect \gls{csi} will leverage the maximum array gain, imperfect \gls{csi} will result in power losses.
Reliable \gls{csi} acquisition will be key to making \gls{wpt} with physically large apertures a reality, thus defining future challenges:
\begin{itemize}
    \item Combining \gls{csi} from different sources promises ultra-reliable and efficient operation.
    \item Synchronization and phase-calibration techniques will enable \gls{cjt} with distributed \gls{6g} radio infrastructures.
    \item The integration of sensing will further allow environment-aware infrastructures to (i) jointly learn the \gls{en} device position and map its surroundings, and (ii) predict \gls{csi} in a closed-loop manner.
\end{itemize}

\bibliographystyle{IEEEtran}
\bibliography{IEEEabrv,IEEEabrvCustom,local}

\end{document}